# Filling the Gaps of Polarity

## Implementing Dependent Data and Codata Types with Implicit Arguments

Bohdan Liesnikov[a] , David Binder[b] , and Tim Süberkrüb[c]

a   Delft University of Technology, Netherlands
b   University of Kent, Canterbury, United Kingdom
c   University of Tübingen, Germany

**Abstract**    The expression problem describes a fundamental tradeoff between two types of extensibility: extending a type with new *operations*, such as by pattern matching on an algebraic data type in functional programming, and extending a type with new *constructors*, such as by adding a new object implementing an interface in object-oriented programming. Most dependently typed languages have good support for the former style through *inductive* types, but support for the latter style through *coinductive* types is usually much poorer. Polarity is a language that treats both kinds of types symmetrically and allows the developer to switch between type representations. However, it currently lacks several features expected of a state-of-the-art dependently typed language, such as implicit arguments. The central aim of this paper is to provide an algorithmic type system and inference algorithm for implicit arguments that respect the core symmetry of the language. Our work provides two key contributions: a complete algorithmic description of the type system backing Polarity, and a comprehensive description of a unification algorithm that covers arbitrary inductive and coinductive types. We give rules for reduction semantics, conversion checking, and a unification algorithm for pattern-matching, which are essential for a usable implementation. A work-in-progress implementation of the algorithms in this paper is available at polarity-lang.github.io. We expect that the comprehensive account of the unification algorithm and our design decisions can serve as a blueprint for other dependently typed languages that support inductive and coinductive types symmetrically.

## The Art, Science, and Engineering of Programming

### 1 Introduction

It is not enough to write a good program. A useful program also needs to be maintained, soon outlives its initial specification and requires continuous modification [43, 44]. The core task for a developer is therefore to build systems that are not just correct, but also easy to maintain and extend [45]. Different programming languages and paradigms provide distinct toolsets for this and thereby shape how we model problems. In this context, we focus on functional (FP) and object-oriented programming (OOP) as two major schools of thought. Below, we argue that both styles have different strengths when it comes to extensibility. These different strengths of both paradigms can be combined in one language, and previous work [10] showed this by developing the dependently typed programming language *Polarity*. This paper extends Polarity with implicit arguments and describes the algorithms for type-checking and unification in detail.

Let us illustrate the advantages of the two different paradigms with an example. Below, we show how sets of natural numbers can be implemented in both styles, inspired[1] by Cook [16]. On the left-hand side, Set is defined in FP style as an algebraic data type (ADT). We represent it as a linked list with constructors Nil and Cons and write two definitions Set.insert and Set.contains by pattern-matching. In Polarity, pattern-matching definitions match on an unnamed argument — in this case, of type Set. On the right-hand side, we model the type Set in OOP style by using interfaces [33] which specify its "methods". To be precise, we write a coinductive type [3, 8, 28] and declare two projections (or "destructors" in Polarity) — .insert and .contains. In this style, code that has access to an object of type Set cannot inspect its internal structure — for example, by pattern matching on it — but can invoke any of the available methods. This definition does not constrain the ways of constructing a set, as long as it implements both methods. We again use a list-like implementation with two (co)definitions Nil and Cons — akin to object constructors. Each of them implements the Set interface by defining the methods using copattern matching [3].

```
1   data Set {                              codata Set {
2     Nil,                                    .insert(x: Nat): Set,
3     Cons(x: Nat, s: Set)                    .contains(f: Nat -> Bool): Bool
4   }                                        }
5   def Set.insert(x: Nat): Set {           codef Nil: Set {
6     Nil => Cons(x, Nil),                    .insert(x) => Cons(x, Nil),
7     Cons(y, s) => Cons(x, Cons(y, s)),      .contains(f) => F,
8   }                                        }
9   def Set.contains(f: Nat -> Bool): Bool { codef Cons(x: Nat, s: Set): Set {
10    Nil => F,                               .insert(y) => Cons(x, Cons(y, s)),
11    Cons(x, s) => f.ap(x).or(s.contains(f)), .contains(f) => f.ap(x).or(s.contains(f)),
12  }                                        }
```

How do these two representations differ when it comes to extensibility? The definition of an ADT such as data Set globally fixes the constructors — a set of ways to create instances of a data type. This means that it is easy to add a new definition like is_empty by pattern matching on those constructors. In contrast, an object-oriented representation such as codata Set globally fixes the methods and extending the Set in this representation with a new observation is_empty is therefore far less convenient as it requires changing all codefinitions. On the upside, it is easy to add new objects that implement the interface. For example, we add a Union constructor on the right, while it would require global changes on the left.

---

[1] Translated to Polarity by Zabarauskas [76].

```
1  def Set.is_empty: Bool {          codef Union(s1 s2: Set): Set {
2    Nil => T,                          .insert(i) => Cons(i, Union(s1, s2)),
3    Cons(x, s) => F,                   .contains(i) => s1.contains(i).or(s2.contains(i)),
4  }                                 }
```

More generally, extending the constructors of a data type means adjusting all definitions that pattern match on it. While for a codata type, extending the projections means adjusting all of the constructors, which are defined by copattern matching. This fundamental extensibility trade-off is known as the expression problem [74]. In this context, many researchers have come to realize the importance of *polarity* [26, 27, 77]: most types can be described as either positive types (data, as on the left) or negative types (codata, as on the right) [29, 38].

In a dependently-typed language, where it is possible to prove properties of the programs and pattern-matching is used extensively, the problem is even more prominent. Unfortunately, most of these languages have not been designed for the ground up with polarity in mind. The need to properly support codata types has been slowly recognized over time [3, 60, 61], but this led to languages where codata types are underdeveloped, a late addition or an afterthought. In order to show what a dependently-typed programming language with symmetric, dependent data and codata types can look like, we presented a language called *Polarity* in a prior publication [10].

The symmetry means that we can freely move between the left and the right side in the example above! This is done algorithmically with global program transformations called defunctionalization [21, 39] and refunctionalization [20]. The developer can thus move between representations to work with the more convenient one.

Motivated by these transformations, prior work [10] also introduces *self parameters* — a way for the types of codata destructors to refer to the object itself. For instance, we can state that any set from the running example becomes non-empty after calling insert on it. Below, we define this property in Polarity on the left and, for comparison, in Agda on the right. Self parameters make codata more expressive than dependent coinductive types in other languages, such as Agda or Rocq, as we explain further in Section 2.2 and Appendix C. Even the current example is slightly awkward on the Agda side since we do not have access to the is_empty *method* inside the definition of Set, only to the *value* of the is_empty field.

```
1   codata Set {                      mutual
2     .insert(x: Nat): Set,             record Set : Type where
3     .contains(f: Nat -> Bool): Bool,    coinductive
4     .is_empty: Bool,                    field
5     (self: Set).insert_non_empty(x: Nat)   insert : (x : Nat) → Set
6       : Eq(self.insert(x).is_empty, F),    contains : (f : Nat → Bool) → Bool
7   }                                        is_empty : Bool
8                                            insert_non_empty : (x : Nat)
9                                                    → is_empty' (insert x) ≡ false
10                                        is_empty' : Set → Bool
11                                        is_empty' s = s .Set.is_empty
```

To make Polarity usable in practice, we would like to omit some parts of the program — for example, type parameters. The Set type shown above was specialized to natural numbers, but generalizes to sets of arbitrary types. On the left in the example below, we add an index[2] for the type of set elements in the data declaration — codata [72] would be similar.

```
1  data Set(_: Type) {
2    Nil(a: Type): Set(a),
3    Cons(a: Type, x: a, s: Set(a)): Set(a),
4  }
5  let example: Set(Nat) {
6    Cons(Nat, 1, Cons(Nat, 2, Cons(Nat, 3, Nil(Nat))))
7  }
```

```
data Set(_: Type) {
  Nil(implicit a: Type): Set(a),
  Cons(implicit a: Type, x: a, s: Set(a)): Set(a),
}
let example: Set(Nat) {
  Cons(1, Cons(2, Cons(3), Nil)))
}
```

However, indices have to be provided to each constructor, making the program verbose. To solve this problem, we add implicit arguments to the language. By marking the relevant inputs as implicit on the use site, we can omit them from the use site. Inferring these implicit arguments automatically requires an implementation of a unification algorithm [2, 53, 79]. Explaining how it works precisely in the presence of the symmetry constraints and self parameters is one of the main goals of this paper.

## 1.1 Contributions

This paper describes a type-checker[3] for a dependently-typed programming language with user-defined dependent data and codata types. Constructors and destructors may take implicit arguments that can be omitted at the use site. The design of this dependently-typed language builds on a prior publication [10], which motivates key aspects of the language by two global transformations: de- and re-functionalization in a dependently-typed setting. Following this prior work, the language does not presuppose the existence of any built-in types — in particular, the type of functions can be defined as a codata type [8, 10].

We extend the system described in our previous paper [10] in multiple directions: Our first core contribution is the presentation of a bidirectional type inference algorithm for the language. We give a complete description of the algorithmic rules, which are essential for a usable implementation. This includes the reduction semantics, conversion algorithm, and index unification, as used for pattern-matching. Our second core contribution is a description of a unification algorithm that covers data and codata types without eta-equality and, essentially, respects the symmetry of the language. To the best of our knowledge, this is the first description of a unification algorithm that covers arbitrary codata types.

**Limitations** The goal of the paper is to provide a clear, comprehensive description of the algorithms, not to formalize the calculus. In fact, the language does not yet have a suitable termination measure for definitions or a positivity criterion for types. As such, we do not prove any meta-theoretical properties about the calculus. As for the type checker: we avoid known sources of non-termination and conjecture that typechecking will terminate given a terminating input program. Similarly, we hypothesize that unification terminates if the terms in the input problems are terminating.

---

[2] We use an index since parameters are not stable under de- and refunctionalization [10].

[3] A work-in-progress compiler implementation is available online at polarity-lang.github.io.

## 1.2 Structure

The rest of this article is structured as follows: We start with Section 2 by introducing the language and the new features from the programmer's point of view and how it is desugared. Then we give the formal core syntax of the language in Section 3. Sections 4 and 5 describe a reduction algorithm and a conversion checker for terms without metavariables. Section 6 describes the index unification algorithm that we use for dependent pattern and copattern matching. Using these ingredients, we present a type inference algorithm for the system in Section 7. Finally, we introduce the contextual metavariables for implicit arguments, the unification algorithm to solve them, and we describe how to extend the rest of the system in Section 9. We close with related work in Section 10, and future work and conclusions in Sections 11 and 12.

## 2 Programming in Polarity

In this section we go through several examples to showcase the language and its features. We focus on the aspects that are unusual compared to other dependently-typed languages, since their impact on the reduction, type checker, and unification algorithms constitutes the novelty of the current paper.

### 2.1 Data Types and Pattern Matching

Inductive data types and dependent pattern matching work similarly to other dependently typed languages. Compared to the previous publication [10] which only specified the rules for top-level matches, we introduce rules for *local* pattern matches with motives [50] in this paper. To illustrate this, recall the data type representation of Set from Section 1 and assume we are in a local context with terms s: Set and n: Nat. We can locally prove a property that is similar to insert_non_empty for s and n:

```
1  s.match as self => Eq(self.insert(n).is_empty, F) {
2      Nil => Refl(F),
3      Cons(_, s) => Refl(F),
4  }
```

In each branch, the "target" self is refined to the constructor Nil or Cons, respectively. This allows us to reduce the motive Eq(self.insert(n).is_empty, F), simplifying the proof goal to a reflexive equality.

## 2.2 Self Parameters

In Section 1, we show an example of a codata type with self parameters, but we do not specify how they can be used. Their design follows from the refunctionalization of pattern matching with motives, as described in detail in prior work [10, sec. 2.5].

This is strictly more expressive than codata types without self parameters, as illustrated in Appendix C. However, typechecking a comatch now means that we need to evaluate expressions referencing the self parameter. For example, for codef Cons, we can prove insert_non_empty as follows:

```
1  codata Set {                              codef Cons(x: Nat, s: Set): Set {
2    ...,                                       .insert(y) => Cons(x, Cons(y, s)),
3    (self: Set).insert_non_empty(x: Nat)       .contains(f) => f.ap(x).or(s.contains(f)),
4      : Eq(self.insert(x).is_empty, F)         .is_empty => F,
5  }                                            .insert_non_empty(_) => Refl(F),
6                                             }
```

When typechecking the case for .insert_non_empty(_), we will reduce the return type Eq(self [Cons(x, s)/self] .insert(x).is_empty, F) to Eq(F, F), so that the proof obligation can be dispatched by reflexivity. This impacts the type inference rules for comatches, as described in Section 7.2.

## 2.3 Functions and Lambdas

Polarity does not provide any built-in types, so any type that we want to use has to be introduced by either a data or codata declaration. This includes non-dependent and dependent function types, which are introduced as two separate codata types.

```
1  codata Fun(_, _: Type) {                  codata Pi(a: Type, _: a -> Type) {
2    Fun(a, b).ap(implicit a b: Type, x: a): b,  Pi(a, p).dap(a: Type, p: a -> Type, x: a): p.ap(x),
3  }                                           }
```

Non-dependent functions are defined by a single observation ap on a function of type Fun(a, b) which takes a value x of type a and returns a value of type b. The arguments a and b in Fun(a, b).ap are bound in the parameter list that follows ap. Polarity allows us to introduce the operator a -> b as syntactic sugar for Fun(a, b). We can then define dependent functions Pi(a, p) where the return type is computed by applying the non-dependent function p to the input.

The language also provides syntactic sugar for comatches with a single observation. In the example below, a lambda function is applied to all elements of a list. In the compiler, \ap(_, _, x) => x.even desugars to comatch { .ap(_, _, x) => x.even } .

```
1  let list_1: List(Nat) {Cons(1, Cons(2, Cons(3, Nil)))}
2  let list_2: List(Bool) {list_1.map(\ap(_, _, x) => x.even)}
```

## 2.4 Equality of Matches and Comatches

Along with de- and re-functionalization, Polarity provides [10] another refactoring tool: (co)match lifting. It turns local lambda expressions and other comatches into top-level codefinitions and local matches into top-level definitions. For example, lifting the lambda expression from the example above yields the following program:

```
1  let list_1: List(Nat) {Cons(1, Cons(2, Cons(3, Nil)))}
2  codef EvenFun: Fun(Nat, Bool) { .ap(_, _, x) => x.even }
3  let list_2: List(Bool) {list_1.map(EvenFun)}
```

Lifting is a prerequisite for applying defunctionalization. Defunctionalization creates a data type with a constructor for each top-level codefinition in the original program (in this case just the one function EvenFun). Thereby, if we defunctionalize the Fun type, we get the following program:

```
1  data Fun(a b: Type) { EvenFun: Fun(Nat, Bool) }
2  def Fun(a, b).ap(implicit a b: Type, x: a): b { EvenFun => x.even }
3  let list_1: List(Nat) {Cons(1, Cons(2, Cons(3, Nil)))}
4  let list_2: List(Bool) {list_1.map(EvenFun)}
```

Distinct constructors are not equal, so if distinct comatches are lifted to distinct codefinitions, they can also never be equal? The 2024 paper [10, sec. 4.1] discusses this issue at length, but fundamentally, Polarity preserves judgmental equality through de- and re-functionalization. In practice, this means that every comatch in the source code receives an internal label that is unique in the whole program. Thus, comatches can be equal only if they originate from the same source code position — i.e., have the same label. The conversion checker then compares (Section 5.5) the labels and variable assignments in the closure to determine whether two comatches are judgmentally equal. Similar constraints apply to local pattern matches, as we will see in Sections 5 and 9.

Finally, $\eta$-equality for general codata types is undecidable [9]. If we were to introduce it for a blessed function type to equate f: a -> b and .ap(_, _, y) => f.ap(y), we would face similar challenges during de- and re-functionalization.

For the user this means a different judgmental equality, one where no-confusion [51] for comatches is provable. However, the user can define a new relation to capture the familiar notion of equality by using a setoid [7] or the right definition of bisimilarity [9, 18]. For the implementation this means we do not have to face a known problem and instead deal with symmetric equality rules.

## 3 Formal Syntax

In the previous section we have seen what the language looks like for the programmer; Table 1 describes the formal equivalent that we use in the elaboration and typechecking rules. In this paper we use a named representation for readability and the Barendregt [6] convention to keep variable names fresh, while we use de Bruijn indices [22] in the implementation. We write $X, \ldots$ instead of the more verbose $X_1, \ldots, X_n$ and use explicit quantification $\forall i : [\ldots X_i \ldots]$ whenever we quantify over all elements of such a list in a rule. We use $T$ for names of data and codata types such as List, Fun and Stream, $K$ for names of data constructors such as True, Nil and Cons, and $D$ for destructors such as ap, hd and tl. Lowercase Latin letters $x$, $y$, $z$ are used for both term and type variables.

Terms consist of variables $x$, type-annotated terms $e : t$ which change the typechecker mode from inference to checking, non-recursive let-bindings **let** $x : t := e; e$

■ **Table 1** Syntax without metavariables

| | | | |
|---|---|---|---|
| $T \in \text{TypeNames}$ | $K \in \text{Constructors}$ | $D \in \text{Destructors}$ | $x, y, z \in \text{Vars}$ |
| $e, s, t$ | ::= | $x$ | *Variable* |
| | \| | $e : t$ | *Type annotation* |
| | \| | **let** $x : t := e; e$ | *Let-binding* |
| | \| | **Type** | *Universe* |
| | \| | $T\sigma$ | *Type constructor* |
| | \| | $K\sigma$ | *Data constructor* |
| | \| | $e.\textbf{match}\ x\ \sigma\ \textbf{as}\ z\ \textbf{return}\ s\ \{a, \dots\}$ | *Match* |
| | \| | $e.D\sigma$ | *Destructor* |
| | \| | $\textbf{comatch}\ x\ \sigma\ \{o, \dots\}$ | *Comatch* |
| $a$ | ::= | $K\Xi \mapsto e$ | *Case* |
| | \| | $K\Xi\ \textbf{absurd}$ | *Absurd case* |
| $o$ | ::= | $D\Xi \mapsto e$ | *Cocase* |
| | \| | $D\Xi\ \textbf{absurd}$ | *Absurd cocase* |
| $\Gamma$ | ::= | $()\ \|\ \Gamma, x : t\ \|\ \Gamma, x : t := e$ | *Typing context* |
| $\Psi, \Xi$ | ::= | $()\ \|\ x : t, \Xi$ | *Telescope* |
| $\rho, \sigma$ | ::= | $()\ \|\ (e, \sigma)$ | *Telescope substitutions* |
| $\delta$ | ::= | $\textbf{data}\ T\Psi\ \{\ K\Xi : T\rho, \dots\ \}$ | *Data declaration* |
| | \| | $\textbf{codata}\ T\Psi\ \{\ (x : T\rho).D\Xi : t, \dots\ \}$ | *Codata declaration* |
| $\Theta$ | ::= | $()\ \|\ \Theta, \delta$ | *Global environment* |

and the impredicative type universe **Type**. Impredicative here means that we have a typing rule that says that **Type** is an inhabitant of itself. We discuss the consequences of this choice in Section 7.

Elements of user-defined data types are introduced by $K\sigma$, i.e., the constructor $K$ applied to the list of arguments $\sigma$, and eliminated using a pattern matching expression $e.\textbf{match}\ x\ \sigma\ \textbf{as}\ z\ \textbf{return}\ s\ \{a, \dots\}$. This expression matches the scrutinee $e$ against a list of cases $a_1$ to $a_n$ that are headed by a constructor $K$, which binds variables in the telescope $\Xi$. A case either has an expression on its right-hand side or is marked as **absurd**; type information ensures that the scrutinee can never match against the constructor mentioned in these absurd cases. The return type of the pattern matching expression is annotated as $s$ and can refer to the scrutinee, which is bound to the name $z$ by the "**as** $z$" part of the expression. Each match has a globally unique name $x$ that is needed for refunctionalization. During desugaring, we computed the free variables of the match body and store these in an identity substitution $\sigma$, which we attach to the match in the internal representation. Going forward, we will omit parts of the match expression when they are not relevant for the text.

Elements of user-defined codata types are introduced using comatch expressions, written **comatch** $x$ $\sigma$ $\{o, \ldots\}$ and eliminated using destructor applications $e.D\sigma$. Comatches can use the name $x$ to refer to themselves in the right-hand sides of their cocases. A similar feature for matches would be a recursive **let**, but we do not discuss this here since the rules are quite verbose. The body of a comatch consists of a list of cocases $o$ which specify the body $e$ that is returned on a method invocation. If the typesystem ensures that a method can never be called, we use the **absurd** cocase instead. As with pattern matches, desugaring attaches a globally unique name $x$ and substitution $\sigma$, which closes over the free variables of the comatch body. For example, the expression comatch { .ap(y) => y } desugars to **comatch** $x$ $\mathrm{id}_{\mathrm{FV}}$ { ap($y$) $\mapsto$ $y$ }.

Both typing contexts $\Gamma, \Delta$ and telescopes $\Psi, \Xi$ store the types of variables, but they are used for slightly different purposes and must be distinguished. Typing contexts provide information about the environment in which an expression $e$ is typechecked. They therefore have to be closed, which means that all free variables within a context are bound at some position to the left, and they are also defined as snoc-lists so that we can extend them on the right when we go under a binder during typechecking. Contexts can also store the term that a variable is bound to, which is needed during typechecking to store locally-bound variables that are inlined and normalized on demand. Telescopes [23], on the other hand, are not used to provide the typing context of an expression $e$. Instead, they specify the types of arguments of a constructor $K$ or destructor $D$, and are used to extend a typing context $\Gamma$ when we pattern match on $K\Xi$ or to check that the arguments in an application $K\sigma$ have the correct type. Both of these applications require telescopes to be cons-lists that grow to the left. It also does not make sense, for these use cases, to store a term $e$ with variable binding, as we do in typing contexts. A substitution $\sigma$ is a list of expressions $e$ and can be typed against a telescope. We do not allow partial application and use substitutions as complete lists of arguments to type constructors $T$, data constructors $K$, and destructors $D$.

Further in the paper we will apply substitutions to terms and other substitutions, written as $t[\sigma]$ or $\rho[\sigma/\Xi]$. For the purposes of this paper a simple capture-avoiding definition suffices. As will be explained in the later Sections 4 and 7, the only noteworthy detail is that a substitution does not have to go under the binders of a (co)match and instead is applied to the closures they carry.

The global environment $\Theta$ consists of data declarations **data** $T\Psi$ { $K\Xi : T\rho, \ldots$ } and codata declarations **codata** $T\Psi$ { $(x : T\rho).D\Xi : t, \ldots$ }. A data declaration introduces a type constructor $T$ with indices $\Psi$ and a list of constructors $K$, each of which takes arguments specified by $\Xi$ and returns a specific instance $T\rho$ of the data type. A codata declaration introduces a type constructor $T$ with indices $\Psi$ and a list of destructors $D$. Each destructor $(x : T\rho).D\Xi : t$ takes a list of arguments $\Xi$, and can be called on an instance where the indices $\Psi$ have been instantiated to $\rho$. The arguments $\rho$ can use variables bound in $\Xi$, and the return type $t$ of the destructor can additionally use the self parameter $x$. As mentioned in Section 1 and prior work [10], we do not distinguish between parameters and indices, as the other systems, and treat all arguments to type constructors as indices.

## 4 Normal Forms and Reduction

In dependently typed languages we have to interleave typechecking and reduction. In this section we specify an environment-based reduction to weak head normal form (WHNF) using a call-by-name strategy. Definition 1 specifies the syntax of environments, WHNFs and neutral terms, which are terms that are stuck on a variable that blocks further computation. The self types of our system introduce a complication: we have to reduce terms that are potentially ill-typed and non-terminating. This implies that reduction might not terminate and that we have to take care of absurd cases.

**Definition 1** (Neutral Terms, Weak Head Normal Forms and Environments)**.**

| | | |
|---|---|---|
| $n$ | $::=$ | $x \mid n.D\sigma \mid n.\textbf{match } x \; \sigma \textbf{ as } z \textbf{ return } s \; \{a, \dots\}$ Neutral term |
| $w$ | $::=$ | $n \mid \textbf{\textit{Type}} \mid T\sigma \mid K\sigma \mid \textbf{comatch } x \; \sigma \; \{o, \dots\}$ Weak Head Normal Form |
| $\phi$ | $::=$ | $\Gamma \mid \phi, x \mapsto e \mid \phi, x \mapsto_c e$ Local environment |

The environment $\phi$ that we use to make reduction more efficient stores variable bindings $x \mapsto e$ and comatch bindings $x \mapsto_c e$. Environments also store the typing context $\Gamma$ from which reduction was called, because this context can contain the bodies of let-bound variables. This context $\Gamma$ is a prefix of $\phi$ and is never modified or interleaved with the other bindings.

In the table below we define small-step reduction as $e[\phi] \triangleright e'[\phi']$. A variable $x$ that we encounter during reduction can be bound in one of two ways. If it is bound in the environment part then it always has a body and we inline it using the rule ($\textsc{Env}_1$). A variable bound in the typing context $\Gamma$ might have a body or not, but if it has a body then we inline it using the rule ($\textsc{Env}_2$). In the rest of the text we will use $[x \mapsto e] \in \phi$ as a shorthand for these two rules. The rules for looking up comatch bindings $x \mapsto_c e$ are discussed with the reduction of comatches.

$$x[\phi] \quad \triangleright \quad e[\phi] \quad \text{if} \quad [x \mapsto e] \in \phi \qquad\qquad \textsc{Env}_1$$

$$x[\Gamma, \phi] \quad \triangleright \quad e[\Gamma, \phi] \qquad\qquad \textsc{Env}_2$$
$$\text{if} \quad (x : t := e) \in \Gamma, \text{ for some } t$$

$$(K\sigma.\textbf{match } x \; \rho \; \{K\Xi \mapsto e, \dots\})[\phi] \quad \triangleright \quad e[\phi, \rho, \sigma] \qquad\qquad \textsc{Match}$$

$$(\textbf{comatch } x \; \rho \; \{\dots\}.D\sigma)[\phi] \quad \triangleright \quad x.D\sigma[\phi, x \mapsto_c \textbf{comatch } x \; \rho \; \{\dots\}]$$
$$\textsc{Comatch}_1$$

$$(x.D\sigma)[\phi] \quad \triangleright \quad s[\phi, \rho, \sigma] \qquad\qquad \textsc{Comatch}_2$$
$$\text{if} \quad [x \mapsto_c \textbf{comatch } x \; \rho \; \{D\Xi \mapsto s, \dots\}] \in \phi$$

$$(e.\textbf{match } x \; \sigma \; \{a, \dots\})[\phi] \quad \triangleright \quad (e'.\textbf{match } x \; \sigma \; \{a, \dots\})[\phi] \quad \text{if} \quad e \triangleright e' \quad \textsc{Cng}_1$$

$$(e.D\sigma)[\phi] \quad \triangleright \quad (e'.D\sigma)[\phi] \quad \text{if} \quad e \triangleright e' \qquad \textsc{Cng}_2$$

$$(e : t)[\phi] \quad \triangleright \quad (e' : t)[\phi] \quad \text{if} \quad e \triangleright e' \qquad \textsc{Cng}_3$$

$$(\textbf{let } x := s; e)[\phi] \quad \triangleright \quad e[\phi, x \mapsto s] \qquad\qquad \textsc{Let}$$

The $\textsc{Match}$ rule for data types fires when a constructor $K\sigma$ meets a pattern match. During evaluation, we must give the branches access to the constructor arguments $\sigma$ as well as any bindings contained in the closure $\rho$. We do this by converting both $\sigma$ and $\rho$ to a sequence of let-bound environment variables when we write $[\phi, \rho, \sigma]$. We store the arguments without evaluation, following the call-by-name semantics.

Dually, we reduce a comatch when a destructor $D\sigma$ is called on it. In our system, comatches carry unique names $x$ which recursively refer to themselves in its body, and we use the following two rules to reduce them. The COMATCH$_1$ rule replaces the (potentially recursive) comatch by a variable which we add to the environment. This rule is used, for example, in the reduction of Cons from Section 2.2 or an infinite constant stream codefinition. We can then access the recursive binding of $x$ using the COMATCH$_2$ rule. Absurd clauses for both matches and comatches should never be reduced and if we reach them, the input was ill-typed.

There are two congruence rules CNG$_1$ and CNG$_2$ for pattern and copattern matching and we can reduce under type annotations using the rule CNG$_3$. In the remainder of this paper we implicitly assume that we can strip these annotations when necessary, usually to inspect the head of the expression. Finally, let-bound definitions are added to the environment using LET.

We use $\triangleright^*$ for the transitive closure over $\triangleright$ and $e[\phi] \triangleright^* w[\phi']$ if reduction terminates and yields a value in weak-head normal form. Crucially, it produces a value $w$ that may contain variables bound in the environment $\phi'$. We will re-use the environment for conversion checking in Section 5.

We also need to call the reduction from the typechecker, for example, to check that a comatch has appropriate type $T$. We define $\Gamma \vdash e \rightsquigarrow w$ to produce a pure term in WHNF in two steps. First, it reduces $e[\Gamma] \triangleright^* w'[\Gamma, \phi]$, then quotes back $w'$ with an environment $\phi$ to produce a term $w$ scoped over $\Gamma$. In order to quote a term $w'$ we traverse it and substitute any variables bound in the environment $\phi$, but not $\Gamma$. However, the substitution does not go under matches and comatches: Given **comatch** $x \, \sigma \, \{ \ldots \}$ we substitute values into $\sigma$ but not into the cocases, as shown in the following example:

**let** $y := 42$; **comatch** $x \, (y) \, \{ \, \mathrm{ap}(z) \mapsto y \, \} \rightsquigarrow$ **comatch** $x \, (42) \, \{ \, \mathrm{ap}(z) \mapsto y \, \}$

## 5   Conversion Checking

Conversion checking $\mathrm{conv}(\Gamma, e_1, e_2)$ is invoked by the rule EMBED of Section 7 to decide if the two terms $e_1$ and $e_2$ are definitionally equal in the typing context $\Gamma$. As conversion is decidable for data [17] and, without $\eta$-equality, for codata [9], it is enough to distinguish two outcomes:

**Positive Decision** The algorithm determines that the terms are definitionally equal. In that case, we write $\mathrm{conv}(\Gamma, e_1, e_2) = \mathbf{ok}$.

**Negative Decision** The algorithm determines that the terms cannot possibly be equal. We write $\mathrm{conv}(\Gamma, e_1, e_2) = \mathbf{conflict}$ to indicate this outcome.

After adding metavariables to the system, conversion checking will be part of the unification algorithm, defined in Section 9. For the system without metavariables we give a simpler, self-contained definition here.

**Filling the Gaps of Polarity**

The internal state of the algorithm (Definition 2) consists of constraints $Cs$ which are tracked as part of the unification context $\mathscr{U}$. Each constraint carries a reduction environment $\phi$, defined in Table 1. When the conversion checker is called from the typechecker as $\mathrm{conv}(\Gamma, e_1, e_2)$, we initialize the context $\mathscr{U}_1$ to $\{C = \Gamma \vdash e_1 \sim e_2\}$. In the rules below, we will sometimes leave out the environment if it is unchanged, and assume that for every rule which operates on a constraint $e_1 \sim e_2$ there is a symmetric rule which operates on $e_2 \sim e_1$.

**Definition 2** (Conversion checking state).

$$
\begin{array}{llll}
Cs & ::= & [] \quad | \quad (\phi \vdash e_1 \sim e_2) :: Cs \quad | \quad (\phi \vdash \sigma_1 \sim_{arg} \sigma_2) :: Cs & \text{Constraints} \\
\mathscr{U} & ::= & \{C = Cs\} & \text{Unification context}
\end{array}
$$

The unification algorithm performs a series of steps $\mathscr{U}_1 \rightsquigarrow \mathscr{U}_2 \rightsquigarrow \ldots \rightsquigarrow \mathscr{U}_n$. Each step removes the first constraint from the context and decides which rule to apply. If there is an applicable rule, it may add new constraints to the context. We return **ok** if all constraints could be solved, written $\mathscr{U}_1 \rightsquigarrow \ldots \rightsquigarrow \{C = []\}$. If we find conflicting constraints along the way $\mathscr{U}_1 \rightsquigarrow \ldots \rightsquigarrow \bot$, we abort with **conflict**. We do not list all conflict rules here, but the complete set of rules covers all possibilities: if the inputs are terminating, a decision can always be made. This will change once we add metavariables to the system in Section 9. If the input contains a diverging expression, the conversion checker may not terminate either.

While the decision the algorithm arrives at does not depend on the order in which we apply the rules, the efficiency of the implementation does. For example, we first check two sides for $\alpha$-equivalence using the shortcut rule Conv-Alpha before applying the reduction rule Conv-Red.

### 5.1 Conversion Checking for Arguments

Conversion checking for terms $e_1 \sim e_2$ is defined mutually recursive with conversion checking for arguments $\sigma_1 \sim_{arg} \sigma_2$ which iterates over the two argument lists and checks that each pair is convertible.

$$
\begin{array}{lll}
\{C = () \sim_{arg} () :: Cs\} & \rightsquigarrow & \{C = Cs\} \qquad\qquad\qquad\qquad\qquad\quad \text{C-Nil} \\
\{C = (e_1, \sigma_1) \sim_{arg} (e_2, \sigma_2) :: Cs\} & \rightsquigarrow & \{C = e_1 \sim e_2 :: \sigma_1 \sim_{arg} \sigma_2 :: Cs\} \quad \text{C-Cons}
\end{array}
$$

### 5.2 Shortcut and Variables

An important optimization is to avoid expensive computations by always checking terms for $\alpha$-equality first, which we write as $e_1 \simeq e_2$.

$$
\{C = e_1 \sim e_2 :: Cs\} \quad \rightsquigarrow \quad \{C = Cs\} \quad \text{if } e_1 \simeq e_2 \quad \text{Conv-alpha}
$$

The shortcut rule also handles constraints such as $x \sim x$ and $\mathbf{Type} \sim \mathbf{Type}$, so there is no need for separate rules for atoms. However, for these atoms, we do want additional negative rules when $\alpha$-equality fails, for example:

$$
\{C = w \sim \mathbf{Type} :: Cs\} \quad \rightsquigarrow \quad \bot \quad \text{if } w \not\simeq \mathbf{Type} \quad \text{Conv-Type-Bot}
$$

$$
\{C = x \sim y :: Cs\} \quad \rightsquigarrow \quad \bot \quad \text{if } x \not\simeq y \qquad \text{Conv-Var-Bot}
$$

### 5.3 Reduction

If the terms are not $\alpha$-equal, we have to reduce them to WHNF to make further progress, since knowing the heads of the terms is usually enough to know which rule to apply. The reduction algorithm presented in Section 4 tells us how to do this. This rule is used, for example, when type checking Refl against the *reduced* return type in Section 2.2.

$$\{C = \phi_1 \vdash e_1 \sim e_2 :: Cs\} \quad \rightsquigarrow \quad \{C = \phi_3 \vdash w_1 \sim w_2 :: Cs\} \quad \textsc{Conv-red}$$
$$\text{where} \quad e_1[\phi_1] \triangleright^* w_1[\phi_2]$$
$$e_2[\phi_2] \triangleright^* w_2[\phi_3]$$

The focused constraint $e_1 \sim e_2$ already comes with its reduction environment $\phi_1$. In this environment, we first reduce $e_1$ to a value $w_1$ in WHNF that lives in a new environment $\phi_2$. We then reuse this environment to reduce $e_2$ to $w_2$, producing a third environment $\phi_3$. While doing so, we perform an implicit weakening of $e_2$ from environment $\phi_1$ to $\phi_2$, and a weakening of $w_1$ from environment $\phi_2$ to $\phi_3$. Both of these operations are to make sure that let-bound variables introduced by one reduction sequence are not used in the other one.

The Conv-red rule is the only one that requires tracking of the reduction environment during conversion. This is done to preserve shared let-bound terms on both sides of the constraint. One can also track the environment for each side of the conversion problem separately, as long as the context prefix is shared. Alternatively, we can inline the let-bound variables from the environments by switching from $\triangleright^*$ to $\rightsquigarrow$ in the rule above. Both of these option remove the need for weakening, should it be too expensive to perform.

### 5.4 Constructors and Destructors

Type and data constructors are injective and enjoy non-confusion. Injectivity — for type constructors, but the same holds for data constructors — means that $T\sigma$ can only be equal to $T\sigma'$ if $\sigma$ is equal to $\sigma'$. No-confusion means that $T\sigma$ and $T'\sigma'$ can never be equal if $T$ is not equal to $T'$, whatever $\sigma$ and $\sigma'$ might be.

$$\{C = T\sigma \sim T\rho :: Cs\} \quad \rightsquigarrow \quad \{C = \sigma \sim_{\text{arg}} \rho :: Cs\} \quad \textsc{Conv-TCtor}$$
$$\{C = K\sigma \sim K\rho :: Cs\} \quad \rightsquigarrow \quad \{C = \sigma \sim_{\text{arg}} \rho :: Cs\} \quad \textsc{Conv-DCtor}$$

$$\{C = T_1\sigma \sim T_2\rho :: Cs\} \quad \rightsquigarrow \quad \bot \quad \text{if } T_1 \not\simeq T_2 \quad \textsc{Conv-TCtor-Bot}$$
$$\{C = K_1\sigma \sim K_2\rho :: Cs\} \quad \rightsquigarrow \quad \bot \quad \text{if } K_1 \not\simeq K_2 \quad \textsc{Conv-DCtor-Bot}$$

We can decompose a constraint between destructor invocations on two neutrals $n_1$ and $n_2$ if the destructors are the same. Finally, different constructors are never equal.

$$\{C = n_1.D \ \sigma_1 \sim n_2.D \ \sigma_2 :: Cs\} \quad \rightsquigarrow \quad \{C = n_1 \sim n_2 :: \sigma_1 \sim_{\text{arg}} \sigma_2 :: Cs\} \quad \textsc{Conv-Dtor}$$
$$\{C = n_1.D_1\sigma_1 \sim n_2.D_2\sigma_2 :: Cs\} \quad \rightsquigarrow \quad \bot \quad \text{if} \quad D_1 \not\simeq D_2 \quad \textsc{Conv-Dtor-Bot}$$

## 5.5 Equality for (Co)matches

As explained in Section 2, comatches have an unusual notion of equality, which does not include $\eta$. Besides Conv-alpha, comatches enjoy the following two rules.

$$\{C = \textbf{comatch } x_1\ \sigma_1\{\ldots\} \sim \textbf{comatch } x_2\ \sigma_2\{\ldots\} :: Cs\} \rightsquigarrow \qquad \textsc{Conv-Comatch}$$
$$\{C = \sigma_1 \sim \sigma_2 :: Cs\} \qquad \text{if } x_1 = x_2$$

$$\{C = \textbf{comatch } x_1\ \sigma_1\{\ldots\} \sim \textbf{comatch } x_2\ \sigma_2\{\ldots\} :: Cs\} \rightsquigarrow \quad \textsc{Conv-Comatch-bot}$$
$$\bot \qquad \text{if } \quad x_1 \neq x_2$$

Following Conv-Comatch, two comatches are equal only if their labels and closures are equal. The inequality rule Conv-Comatch-bot is particularly unusual and allows us to derive no-confusion for comatches. This is motivated by the defunctionalization program transformation, which will turn any comatch into a constructor with label $x$ and argument list $\sigma$.

Dually, the refunctionalization transformation will turn any pattern match into a destructor, which leads to similar conversion rules. An implementation not concerned with de- and refunctionalization could leave these out.

$$\{C = e_1.\textbf{match } x_1\ \sigma_1\{\ldots\} \sim e_2.\textbf{match } x_2\ \sigma_2\{\ldots\} :: Cs\} \rightsquigarrow \qquad \textsc{Conv-Match}$$
$$\{C = e_1 \sim e_2 :: \sigma_1 \sim \sigma_2 :: Cs\} \qquad \text{if } x_1 = x_2$$

$$\{C = e_1.\textbf{match } x_1\ \sigma_1\{\ldots\} \sim e_2.\textbf{match } x_2\ \sigma_2\{\ldots\} :: Cs\} \rightsquigarrow \quad \textsc{Conv-Match-bot}$$
$$\bot \qquad \text{if } \quad x_1 \neq x_2$$

## 6  Index Unification

In this section, we introduce the index unification algorithm that is used for dependent pattern matching. This algorithm follows the design of Cockx, Devriese, and Piessens [15], which can be consulted as a reference for the design constraints and the meta theory of index unification. Note that the index unification algorithm ($\texttt{unify}_\iota$) which is described in this section is different from the unification algorithm ($\texttt{conv}$) described in Sections 5 and 9. The reason for having two different algorithms is that they serve different purposes. Index unification is used to locally refine information about the return type and the typing context that we gain by (co)pattern matching on an indexed (co)data type. Unification, on the other hand, is used for checking judgmental equality of terms and finding solutions for metavariables. Index unification never solves metavariables and has to give up if it encounters an unsolved one.

The typechecker invokes $\texttt{unify}_\iota^{\text{args}}(\sigma_1, \sigma_2)$ to perform index unification on the two substitutions $\sigma_1$ and $\sigma_2$, which results in one of the following outcomes:

**Positive Decision** The algorithm terminates and finds a most general unifier of the unification problem. In that case, we write $\texttt{unify}_\iota^{\text{args}}(\sigma_1, \sigma_2) = \theta$.

**Negative Decision** The algorithm terminates and determines that a unifier cannot possibly exist. We write $\texttt{unify}_\iota^{\text{args}}(\sigma_1, \sigma_2) = \textbf{conflict}$ to indicate this outcome.

**Failure** The algorithm could neither determine that a most general unifier exists, nor that there cannot be a unifier. We write $\texttt{unify}_\iota^{\text{args}}(\sigma_1, \sigma_2) = \textbf{fail}$ and report to the user that the program could not be typechecked successfully.

■ **Table 2** Context substitutions

| | | |
|---|---|---|
| $\theta, \eta \quad ::= \quad () \mid \theta, x \mapsto e$ | | *Context substitutions* |

Index unification repeatedly reduces both sides of the equation to WHNF and applies the following rules until no more constraints remain or a contradiction has been found. If none of the rules below apply to the remaining constraints, then we give up and return **fail** as described above.

**Arguments** These rules lift index unification from terms to substitutions.

$$\frac{}{\mathtt{unify}_\iota^{\mathrm{args}}((),()) = ()} \text{ Nil-Nil}$$

$$\frac{\mathtt{unify}_\iota(e_1, e_2) = \theta_1 \qquad \mathtt{unify}_\iota^{\mathrm{args}}(\sigma_1\,\theta_1, \sigma_2\,\theta_1) = \theta_2}{\mathtt{unify}_\iota^{\mathrm{args}}(e_1 :: \sigma_1, e_2 :: \sigma_2) = \theta_2} \text{ Cons-Cons}$$

**Solution and Cycle** These rules solve a constraint where one side consists of a bound variable.

$$\frac{x \notin \mathrm{FV}(t)}{\mathtt{unify}_\iota(x, t) = x \mapsto t} \text{ Solution}_1 \qquad \frac{x \notin \mathrm{FV}(t)}{\mathtt{unify}_\iota(t, x) = x \mapsto t} \text{ Solution}_2$$

$$\frac{x \in \mathrm{FV}(t)}{\mathtt{unify}_\iota(x, t) = \mathbf{conflict}} \text{ Cycle}_1 \qquad \frac{x \in \mathrm{FV}(t)}{\mathtt{unify}_\iota(t, x) = \mathbf{conflict}} \text{ Cycle}_2$$

**Deletion** This rule solves a constraint between two identical terms. This rule implies axiom K, so it cannot be used if this is undesired [14].

$$\frac{}{\mathtt{unify}_\iota(t, t) = ()} \text{ Deletion}$$

**Injectivity** This rule solves a constraint of the form $T\sigma \equiv T\sigma'$ or $K\sigma \equiv K\sigma'$.

$$\frac{\mathtt{unify}_\iota^{\mathrm{args}}(\sigma_1, \sigma_2) = \theta}{\mathtt{unify}_\iota(K\sigma_1, K\sigma_2) = \theta} \text{ Inj-Ctor} \qquad \frac{\mathtt{unify}_\iota^{\mathrm{args}}(\sigma_1, \sigma_2) = \theta}{\mathtt{unify}_\iota(T\sigma_1, T\sigma_2) = \theta} \text{ Inj-TyCtor}$$

**Conflict** This rule solves a constraint between two different data constructors or type constructors.

$$\frac{K_1 \neq K_2}{\mathtt{unify}_\iota(K_1\sigma_1, K_2\sigma_2) = \mathbf{conflict}} \text{ Conflict}_1$$

$$\frac{T_1 \neq T_2}{\mathtt{unify}_\iota(T_1\sigma_1, T_2\sigma_2) = \mathbf{conflict}} \text{ Conflict}_2$$

$$\frac{}{\mathtt{unify}_\iota(T\sigma_1, K\sigma_2) = \mathbf{conflict}} \text{ Conflict}_3$$

Finally, for compatibility with de-/refunctionalization, we also need to add the following, unusual rules:

$$\frac{x_1 \neq x_2}{\mathtt{unify}_\iota(\mathbf{comatch}\ x_1\ \sigma_1\ \{\dots\}, \mathbf{comatch}\ x_2\ \sigma_2\ \{\dots\}) = \mathbf{conflict}} \text{ Conflict}_4$$

$$\frac{\mathtt{unify}_\iota^{\mathrm{args}}(\sigma_1, \sigma_2) = \theta}{\mathtt{unify}_\iota(\mathbf{comatch}\ x\ \sigma_1\ \{\dots\}, \mathbf{comatch}\ x\ \sigma_2\ \{\dots\}) = \theta} \text{ Inj-Comatch}$$

These rules ensure, for example, that a list can be indexed by its length expressed using a codata representations of natural numbers (see Appendix C).

These rules are one reason why our system is incompatible with $\eta$-laws for codata, as described in the introduction. An implementation not concerned with de- and refunctionalization could leave these rules out.

If none of these rules apply, then $\mathtt{unify}_t(t_1, t_2)$ returns **fail**.

## 7 The Basic Type Inference Algorithm

In this section we combine reduction (Section 4), conversion checking (Section 5), and index unification (Section 6) to define a bidirectional typechecker for our system. For every syntactic construct in Table 1, we introduce both a checking and an inference rule. We write $\Gamma \vdash e \Rightarrow t$ if the type $t$ is *inferred* for the expression $e$, and $\Gamma \vdash e \Leftarrow t$ if we check that the expression $e$ has the type $t$. The rules used for checking whether a program is well-formed can be found in Appendix B.

Telescopes are well-formed when they adhere to the rules Tele-Nil and Tele-Cons, and to check whether the expressions in a substitution can be typed against a telescope, we use the rules Subst-Nil and Subst-Cons.

$$\frac{}{\Gamma \vdash ()\ \textbf{tele}}\ \text{Tele-Nil} \qquad \frac{\Gamma \vdash t \Leftarrow \textbf{Type} \qquad \Gamma, x : t \vdash \Xi\ \textbf{tele}}{\Gamma \vdash x : t, \Xi\ \textbf{tele}}\ \text{Tele-Cons}$$

$$\frac{}{\Gamma \vdash () \Leftarrow ()}\ \text{Subst-Nil} \qquad \frac{\Gamma \vdash e \Leftarrow t \qquad \Gamma \vdash \sigma \Leftarrow \Xi[e/x]}{\Gamma \vdash (e, \sigma) \Leftarrow x : t, \Xi}\ \text{Subst-Cons}$$

Going forward, we will not specify the checking rule when there is a corresponding inference rule. Instead, we add an embedding rule to switch modes from checking to inference. Type annotations allow switching the other way: from inference to checking mode. This uses the conversion checker, defined in Section 5. These two rules have been named Ascribe and Embed by McBride [49].

$$\frac{\Gamma \vdash e \Leftarrow t}{\Gamma \vdash (e : t) \Rightarrow t}\ \text{Ascribe} \qquad \frac{\Gamma \vdash e \Rightarrow t_1 \qquad \mathtt{conv}(\Gamma, t_1, t_2) = \textbf{ok}}{\Gamma \vdash e \Leftarrow t_2}\ \text{Embed}$$

Type-checking **let** bindings can happen both in inference and checking mode, which we propagate to the body of the let. The definition of the let-bound variable is always checked against the annotated type. Mind that in both rules $t_1$ has to change scopes: in the Let$_{\text{chk}}$ it is weakened by $x$, in the Let$_{\text{inf}}$ it is strengthened by it.

$$\frac{\Gamma \vdash e_1 \Leftarrow t \qquad \Gamma, x : t := e_1 \vdash e_2 \Leftarrow t_1}{\Gamma \vdash \textbf{let}\ x : t := e_1; e_2 \Leftarrow t_1}\ \text{Let}_{\text{chk}} \qquad \frac{\Gamma \vdash e_1 \Leftarrow t \qquad \Gamma, x : t := e \vdash e_2 \Rightarrow t_1}{\Gamma \vdash \textbf{let}\ x : t := e_1; e_2 \Rightarrow t_1}\ \text{Let}_{\text{inf}}$$

We look up the type of a variable in the typing context, regardless of whether it has been introduced via a let-binding or as an argument.

$$\frac{(x : t) \in \Gamma}{\Gamma \vdash x \Rightarrow t}\ \text{Var} \qquad \frac{(x : t := e) \in \Gamma}{\Gamma \vdash x \Rightarrow t}\ \text{Var-let}$$

We assume the Type-in-Type axiom, which means that we only have one type universe **Type**, which is contained in itself.

$$\frac{}{\Gamma \vdash \textbf{Type} \Rightarrow \textbf{Type}}\ \text{Univ}$$

This is known to be inconsistent [34, 40], but we also do not currently enforce termination or productivity in our system. As adding the Type-in-Type axiom merely adds another source of divergence [69], we would not gain anything from avoiding this paradox, e.g. by using a hierarchy of universes. We therefore follow Eisenberg [30] and opt for a simpler presentation using the Type-in-Type axiom.

We can infer for any type constructor $T\sigma$ that it lives in the impredicative universe **Type** if we find a corresponding data or codata declaration in the global environment $\Theta$. This data or codata declaration must declare the type constructor with indices $\Psi$, so we check that the substitution $\sigma$ matches these indices.

$$\frac{\textbf{data } T\Psi \ \{\dots\} \in \Theta \quad \Gamma \vdash \sigma \Leftarrow \Psi}{\Gamma \vdash T\sigma \Rightarrow \textbf{Type}} \ \text{Data} \qquad \frac{\textbf{codata } T\Psi \ \{\dots\} \in \Theta \quad \Gamma \vdash \sigma \Leftarrow \Psi}{\Gamma \vdash T\sigma \Rightarrow \textbf{Type}} \ \text{Codata}$$

The rule for constructors infers, for example, the type $\text{Vec}(\text{Z}, \text{Bool})$ for the expression $\text{VNil}(\text{Bool})$. We first look up the definition $K\Xi : T\rho$ of the constructor in its corresponding data declaration. We then check whether the arguments $\sigma$ of the constructor $K$ can be typed against the declared arguments $\Xi$ using the judgment $\Gamma \vdash \sigma \Leftarrow \Xi$. Finally, we return the type $T\rho$ that was declared in the data declaration, but substitute the concrete arguments $\sigma$ for the variables bound in $\Xi$.

$$\frac{\textbf{data } T\Psi \ \{ K\Xi : T\rho, \dots \} \in \Theta \quad \Gamma \vdash \sigma \Leftarrow \Xi}{\Gamma \vdash K\sigma \Rightarrow T\rho[\sigma/\Xi]} \ \text{Ctor}$$

Given a destructor call, such as a function call $e.\text{ap}(x)$, we first look up the destructor in some codata declaration in the program environment. We then ensure that the destructor arguments $\sigma$ are compatible with the declared telescope $\Xi$. Finally, just like in the Ctor rule, indices in the type $T\rho$ of $e$ may depend on the destructor arguments $\sigma$ and have to be substituted, since we support indexed codata types [72] In both Ctor and Dtor, $\rho$ is a list of terms that can be typed with a telescope $\Psi$. It gets mapped by a substitution $\sigma$ from the scope $\Xi$ to $\Gamma$.

$$\frac{\textbf{codata } T\Psi \ \{ (z : T\rho).D\Xi : t, \dots \} \in \Theta \quad \Gamma \vdash \sigma \Leftarrow \Xi \quad \Gamma \vdash e \Leftarrow T\rho[\sigma/\Xi]}{\Gamma \vdash e.D\sigma \Rightarrow t[\sigma/\Xi][e/z]} \ \text{Dtor}$$

### 7.1 Checking Local Pattern Matches

Given a term $e$ of a data type, we can pattern match on it to learn how it was constructed. For instance, if $e$ has type $\text{Vec}(\text{S}(\text{Z}), \text{Bool})$ then we can extract the sole entry $x$ as follows:

$e.\textbf{match} \ \{ \ \text{Cons}(a, n, x, xs) \mapsto x, \text{Nil } \textbf{absurd} \ \}$

The second case is marked as absurd, since we know that the scrutinee $e$ cannot be $\text{Nil}$ based on its type. We can also use pattern matching to prove properties by case analysis. For example, we can prove that negation is involutive:

$b.\textbf{match as } z \textbf{ return } \text{Eq}(\text{Bool}, z.\text{not.not}, z) \ \{ \ \text{T} \mapsto \text{Refl}(\text{Bool}, \text{T}), \ \text{F} \mapsto \text{Refl}(\text{Bool}, \text{F}) \ \}$

Here we use pattern matching with a motive to specify the return type of the match. This return type $\mathtt{Eq}(\mathtt{Bool}, z.\mathtt{not}.\mathtt{not}, z)$ refers to the scrutinee $b$ by the name $z$. To infer the type of a pattern match consider the following MATCH rule:

$$\frac{\begin{array}{ccc} \Gamma \vdash e \Rightarrow s\ \text{❶} & \Gamma \vdash s \rightsquigarrow T\rho : \mathbf{Type}\ \text{❷} & \mathbf{data}\ T\Psi\ \{\ \dots\ \} \in \Theta\ \text{❸} \\ \Gamma, z : T\rho \vdash t_2 \Leftarrow \mathbf{Type}\ \text{❹} & \forall i : [\Gamma \vdash a_i[\sigma] \Leftarrow (z : T\rho).t_2]\ \text{❺} \end{array}}{\Gamma \vdash e.\mathbf{match}\ x\ \sigma\ \mathbf{as}\ z\ \mathbf{return}\ t_2\ \{a, \dots\} \Rightarrow t_2[e/z]}\ \textsc{Match}$$

First, we infer the type $s$ of the scrutinee $e$ ❶. We then ensure that $s$ normalizes to a type constructor $T\rho$ ❷, allowing us to look up the definition of $T$ in the program ❸. If a motive is specified, we check that $t_2$ is a valid return type which can depend on $z$ ❹. We then check all clauses of the match, which we will describe in detail below ❺. Finally, we infer that the type is $t_2$, with $e$ substituted for $z$. As mentioned in Section 3, we carry the substitutions $\sigma$ in the (co)matches to capture free variables that appear inside the clauses. Since the terms inferred or checked during type checking have not been reduced yet, in many cases, $\sigma$ will be an identity substitution on $\Gamma$, and $a_i[\sigma]$ in ❺ is a no-op. However, this is not always the case. The substitution $\sigma$ is not the identity, for example, if it was altered by index unification as part of typechecking an outer (co)match.

Checking the clauses involves two separate rules, CASE-OK and CASE-ABSURD, depending on whether the case has a body or is absurd. These rules produce the judgment $\Gamma \vdash a \Leftarrow (x : T\rho).t$ where $a$ is the case and $(x : T\rho_2).t$ tracks the motive.

$$\frac{\begin{array}{cc} \mathbf{data}\ T\Psi\ \{\ K\Xi : T\rho_1\ \} \in \Theta\ \text{❻} & \mathtt{unify}_\iota(\rho_1, \rho_2) = \theta\ \text{❼} \\ (\Gamma, \Xi)[\theta] \vdash e[\theta] \Leftarrow t[K\mathrm{id}_\Xi/x][\theta]\text{❽} \end{array}}{\Gamma \vdash K\Xi \mapsto e \Leftarrow (x : T\rho_2).\ t}\ \textsc{Case-Ok}$$

$$\frac{\mathbf{data}\ T\Psi\ \{\ K\Xi : T\rho_1\ \} \in \Theta\ \text{❻} \qquad \mathtt{unify}_\iota(\rho_1, \rho_2) = \mathbf{conflict}\ \text{❼}}{\Gamma \vdash K\Xi\ \mathbf{absurd} \Leftarrow (x : T\rho_2).\ t}\ \textsc{Case-Absurd}$$

In both rules, we first look up ❻ the constructor declaration in the program to find out the declared arguments $\rho_1$ and the type constructor $T$. We then use index unification ❼ to compare these declared arguments with the arguments $\rho_2$ that we inferred for the scrutinee. In the implementation, $\rho_1$ and $\rho_2$ exist in different scopes ($\Xi$ and $\Gamma$, respectively) and have to be weakened accordingly. The result of index unification should either be a substitution $\theta$ from $\Gamma, \Xi$ to a unified context in rule CASE-OK or **conflict** in rule CASE-ABSURD. We report an error if unification returns **fail**, if index unification returns a substitution in a case marked as absurd, or if unification returns **conflict** in a case with a right-hand side. If unification succeeds, we check ❽ the body of the clause against the motive with the constructor substituted for $x$, where we substitute $\theta$ into the context, the expression and the type.

There is an important subtlety here: The substitution on contexts, e.g. $\Gamma[\theta]$ does two things. It applies $\theta$ to all types and values in the context but also "marks" any bindings for variables from the domain of $\theta$ by converting them to let-bindings. For example, if $\theta = [a \mapsto \mathtt{Nat}]$ and $\Gamma = (a : \mathsf{Type}, y : a, z : \mathsf{Type} := a)$, then $\Gamma[\theta] = (a : \mathsf{Type} := \mathsf{Nat}, y :$

$\mathsf{Nat}, z : \mathsf{Type} := \mathsf{Nat}$). It is undesirable to keep $a$ unchanged in the context, since the unification algorithm and occurs checker described in Section 9.4.3 use the variables (without bodies) from the context to infer what an implicit argument can depend on. Keeping $a$ available in the environment can lead to unification failure even when a metavariable would be solvable otherwise. Instead of removing $a$ completely, we "mark" them instead, since this preserves more information for better error messages, while keeping occurs-check happy.

### 7.2 Checking Local Copattern Matches

Dually, codata types are introduced using copattern matching as specified by the rule Comatch. We first reduce ❶ the expected type to a type constructor and then proceed to check ❷ all clauses $o$. For checking the clauses, we keep track of the comatch and its type via a variable $x$ bound in the context. This is needed to typecheck a recursive comatch such as the following stream of ones **comatch** ones $\{\mathsf{hd} \mapsto 1, \mathsf{tl} \mapsto \mathsf{ones}\}$, where the two destructors .hd and .tl define the codata type of streams.

$$\frac{\Gamma \vdash t \rightsquigarrow T\rho \; \text{❶} \qquad \forall i : [\Gamma, x : T\rho := \textbf{comatch } x \; \sigma \; \{o, \dots\} \vdash_c o_i] \; \text{❷}}{\Gamma \vdash \textbf{comatch } x \; \sigma \; \{o, \dots\} \Leftarrow t} \; \text{Comatch}$$

When checking individual clauses, we look up ❶ the codata type declaration in the program to get the expected type constructor arguments $\rho_1$. Unification ❷ produces either a $\theta$ or a **conflict**. If it returns a **fail** or does not return the expected result, we report an error, as detailed in Section 7.1. Since each comatch carries a closure-substitution $\sigma$, the body of the comatch in the context is not affected by the unifier $\theta$ in ❷, but the substitution $\sigma$ is. As for $\sigma$ being an identity substitution and the scoping of $\rho$ and $\theta$, the same arguments as in Section 7.1 apply.

$$\frac{\textbf{codata } T\Psi \; \{(z : T\rho_1).D\Xi : t\} \in \Theta \; \text{❶} \qquad \mathtt{unify}_\iota(\rho_1, \rho_2) = \theta \; \text{❷}}{(\Gamma, x := \dots, \Xi)[\theta] \vdash e[\sigma][\theta] \Leftarrow t[x/z][\theta] \; \text{❸}}{\Gamma, x : T\rho_2 := \textbf{comatch } x \; \sigma \; \{o, \dots\} \vdash_c D\Xi \mapsto e} \; \text{Cocase-OK}$$

$$\frac{\textbf{codata } T\Psi\{(z : T\rho_\iota).D\Xi : t\} \in \Theta \; \text{❶} \quad \mathtt{unify}_\iota(\rho_1, \rho_2) = \textbf{conflict} \; \text{❷}}{\Gamma, x : T\rho_2 := \textbf{comatch } x \; \sigma \; \{o, \dots\} \vdash_c D\Xi \; \textbf{absurd}} \; \text{Cocase-Absurd}$$

The rule Cocase-OK is used, for example, to typecheck insert_non_empty on Cons in Section 2.2. Following the Ctor rule, the constructor call Refl(F) infers the type Eq(F, F). Following the Embed rule, this type will then be checked for definitional equality in the context against Eq(self[Cons/self].insert($x$).is_empty, F) in the extended context $x$ : $\mathsf{Nat}, s : \mathsf{Set}, \mathsf{Cons} := \textbf{comatch } \mathsf{Cons}(x, s)\{\dots\}$. The conversion checking algorithm will apply congruence rules and then reduce the left-hand side of the equality constraint to F, yielding a positive decision **ok**. The substitution $\theta$ can be ignored in this example, as Set is not indexed.

<span style="background:blue;color:white;padding:2px 6px;">8</span> **Desugaring Implicit Arguments**

So far we have built a language with data and codata types. However, this language is quite verbose since data and codata types support only indices, which are not inferred by a pure bidirectional type checker. For example, we have to supply both the input and the output type when we apply the identity function to zero: id.ap(Nat, Nat, Z). Implicit arguments allow us to omit them and write id.ap(Z) instead. This section explains how we can add implicit arguments as a surface language feature and how these implicit arguments can be elaborated to contextual metavariables of the intermediate language that the type inference algorithm operates on.

### 8.1 Syntax

We start by extending the syntax (Table 1) with contextual metavariables, following Abel and Pientka [2]. Intuitively, metavariables are placeholder terms that will be filled in later, during unification.

**Definition 3** (Extended Syntax).

$$\alpha, \beta, \gamma \in \textit{Metavariables}$$

$$
\begin{array}{llll}
e, s, t & ::= & \dots & \textit{Table 1} \\
 & | & \alpha[\theta] & \text{Contextual metavariables}
\end{array}
$$

*Contextual* means that each metavariable $\alpha$ comes with a delayed substitution $\theta$, mapping variables from the context that $\alpha$ was created in to the current context, akin to a closure. For example, in the expression **comatch** { ap($x$) ↦ $\alpha[x]$ }.ap(42), a metavariable $\alpha$ is created in the extended context $\Gamma, x : \text{Nat}$. This expression reduces to $\Gamma \vdash \alpha[42]$, which changes the substitution attached to $\alpha$ from $[x \mapsto x]$ to $[x \mapsto 42]$, making the expression well-scoped in just $\Gamma$, without the $x$. We omit the variable, in this case $x$, bound in the substitution when it is unambiguous, writing $[e]$ for $[x \mapsto e]$.

### 8.2 Desugaring and Typing

To typecheck a program such as id.ap(Z) we first have to translate it into the core syntax with metavariables, which is defined in Definition 3. This is handled by the desugaring pass, which inserts a fresh metavariable for every omitted implicit argument. For example, id.ap(Z) will be translated to the core term id.ap($\alpha[], \beta[], Z$). We defer inserting the actual substitution until we have the precise context available during typechecking.

When we encounter such a metavariable in the typechecker, we initialize the delayed substitution $\theta$ with an identity substitution $\text{id}_\Gamma$. For the example above we will get the term id.ap($\alpha[\text{id}_\Gamma], \beta[\text{id}_\Gamma], Z$). The identity substitution does not include terms for variables that have a body in $\Gamma$ — meaning they were either let-bound or refined by index unification. These variables are not free in the metavariable, so it should not depend on them. The second step is to register the metavariable in the map $\Delta$ (Definition 4). After this, we replace $\alpha[]$ by $\alpha[id_\Gamma]$ in the AST.

In the current implementation, we return an error if we encounter a metavariable in inference mode. An alternative would be to generate a fresh metavariable for the type too, but our initial approach results in poor error messages.

**Definition 4** (Metavariable Map)**.**

$$\Delta \quad ::= \quad (\Gamma \vdash \alpha : t) :: \Delta \qquad \text{Unsolved metavariable}$$
$$| \quad (\Gamma \vdash \alpha : t := e) :: \Delta \qquad \text{Solved metavariable}$$
$$| \quad [] \qquad \text{Empty map}$$

## 9   Solving Metavariables

In Section 8 we described how to desugar implicit arguments to metavariables; this section describes how to solve them. We therefore extend the conversion checker of Section 5 so that it can also work on terms containing metavariables and find their solutions. These solutions should be the most general ones, which means that the unification algorithm has to find the most general unifier (MGU) [2, 53]. Finding the most general unifier for a higher-order unification problem is, in general, undecidable. We can therefore provide only a sound under-approximation: If we solve a metavariable then the solution is the most general one, but we cannot guarantee in every case to find a most general solution if it exists.

### 9.1  State

We define unification state to be the conversion state introduced in Definition 2, extended by the metavariable map $\Delta$ from Definition 4. The equations $Cs$ stored in $C$ are the same as in Definition 2 but may now contain metavariables in terms. The unification algorithm assumes that all metavariables have been registered in $\Delta$ and **fail**s if it encounters a metavariable that has not been registered.

**Definition 5** (Unification state)**.**

$$\mathscr{U} \quad ::= \quad \{C = Cs, E = \Delta\} \qquad \text{Unification context}$$

### 9.2  Occurrences and Free Variables

In higher-order unification we have to distinguish between different kinds of occurrences of a variable within a term, such as the occurrence of $x$ in $\mathrm{Succ}(x)$. In particular, we have to distinguish between strongly rigid, weakly rigid, and flexible occurrences. (We will simply say *rigid* if the difference between strongly or weakly rigid does not matter.) A formal definition of these concepts, adopted from Abel and Pientka [2, p.13], is contained in Appendix D. Here we only present them informally with some examples and motivation.

**Strongly rigid occurrences**  are used in the occurs checker to catch equations that would result in an infinite solution. Strongly rigid occurrences are in the argument to an inductive constructor, in the substitution to a comatch closure, or in the transitive closure of both. In the following two examples $\alpha$ occurs rigidly on the right-hand side and we do not attempt to solve them. A constraint $\alpha = K(\alpha)$ would lead to an infinite solution $\alpha = K(K(K\ldots))$. The term $\alpha = \textbf{comatch } y \; (x \mapsto \alpha) \; \{D\Xi \mapsto x\}$ could be accepted, but would require the comatch to be parameterized by the value it "loops" on. We prohibit this to keep the symmetry between data and codata values. Instead we allow only equations like $\alpha = \textbf{comatch } y \; () \; \{D\Xi \mapsto \alpha\}$, with solution $\alpha = \textbf{comatch } y \; () \; \{D\Xi \mapsto y\}$.

**Weakly rigid occurrences** are used in the occurs checker and the pruning procedure to determine which variables can appear in the potential solution. Weakly rigid entries can occur in three kinds of places: In a neutral projection, the scrutinee and the arguments are weakly rigid. In a match, the scrutinee, return type, and the closure are weakly rigid. Finally, a rigid occurrence in any sub-term of a term that is in weakly rigid position is also weakly rigid. Intuitively, these entries capture positions that will not disappear when any metavariable solution is substituted, therefore they will always appear in the potential solution. For example, $x$ and $y$, but not $z$ occur weakly rigidly in $x.D(K(y), \alpha[x \mapsto z])$.

**Flexible occurrences** are used in the pruning procedure to determine which variables can be removed. We say that $s$ occurs flexibly in $t$ if it is a sub-term inside a delayed substitution to any metavariable in $t$ such as $z$ in the term $\alpha[x \mapsto z]$.

We use functions $\mathrm{FV}(e)$ to compute all free variables, $\mathrm{FV}^{\mathrm{srig}}(e)$ for all strongly rigidly occurring variables, $\mathrm{FV}^{\mathrm{rig}}(e)$ for all rigidly occurring variables, and $\mathrm{FVM}(e)$ for all metavariables in an expression $e$.

### 9.3 Reduction of Solved Metavariables

In Section 4, we defined reduction $e[\phi] \; \triangleright \; e'[\phi']$ in an environment $\phi$. We now have to extend this relation with the following rule $\textsc{Env-}\Delta$ which inlines a metavariable if it has a solution in the global metavariable map $\Delta$:

$$\alpha[\theta][\phi] \triangleright e[\phi, \theta] \quad \text{if } (\Gamma' \vdash \alpha := e) \in \Delta \qquad\qquad \textsc{Env-}\Delta$$

Since $e$ is scoped with respect to $\Gamma'$, we can weaken it to the new environment $\phi, \theta$. This keeps $e$ well-scoped, since $\theta$ defines a mapping from $\Gamma'$ to $\phi$. Adding $\theta$ to the environment also postpones the substitution of concrete terms from $\theta$ until it becomes necessary during reduction, and since it is appended to $\phi$, the terms in $\theta$ remain well-scoped. Same as in Section 4 and $\textsc{Conv-red}$ in Section 5, one must be mindful of the weakening and sharing. Inlining all metavariable solutions immediately is easier from an implementation perspective, but the runtime and memory usage might increase, depending on the representation.

### 9.4 Unification

With the preliminaries out of the way, we can now describe the unification algorithm. It includes all the rules from the conversion checker, which we do not repeat here.

It is crucial that the congruence rules like Conv-Dtor keep the condition that the head is a neutral term ($n$) and not just a blocked one ($b$):

**Definition 6** (Blocked Terms)**.**

$$
\begin{array}{llll}
b & ::= & \alpha[\theta] \mid b.D\sigma \mid b.\textbf{match } x \ \sigma \textbf{ as } z \textbf{ return } s \ \{a,\ldots\} & \text{Blocked term} \\
w & ::= & \ldots & \textit{Definition 1} \\
& \mid & b &
\end{array}
$$

A blocked term might compute further when a solution to the meta appears, eliminating the projection completely. This means that equating the arguments could lose some of the solutions, since not all destructors are injective.

Congruence and reduction rules simplify the constraints as a prerequisite to infer the solutions for metavariables. In particular, the solving procedure only looks at constraints of the shape $\alpha[\theta] \sim e$. It processes one constraint at a time, with one of four possible outcomes: a simplified constraint, a failure, postponement, or a solution. We output a simplified constraint if we cannot solve it, but can remove a dependency of a metavariable, therefore making it simpler. We fail if the right-hand side of the equation mentions variables not available to the meta on the left-hand side; no solution exists in that case. We postpone [25, 57, ch.3.3, 62] the constraint if the occurs check outputs a negative decision that is not a hard failure. This means that a solution might exist, but we need to wait for a different metavariable to be solved before we can make progress here. Finally, if we can invert the substitution $\theta$, we apply it to $e$ and obtain a solution to $\alpha$ and remove the constraint as solved.

If the algorithm arrives at a state when none of the rules apply and there are still unsolved constraints present in the system, it signals **fail**.[4]

### 9.4.1 Solving a Single Constraint

In this section we describe the process of solving a single constraint, for which we use four operations: pruning $\texttt{prune}(\Delta, \theta, e)$, unification of identical metavariables $\texttt{same}(\Delta, \alpha, \theta_1, \theta_2)$, occurs check $\texttt{occurs}(\alpha, \theta, e)$, and inversion of delayed substitutions $\texttt{invert}(\alpha, \theta, e)$. We give a high-level description of each individual step in this section, but in practice implement one $\texttt{solve}$ procedure for efficiency [2, p. 21].

Pruning tries to get rid of the variables that the metavariable $\alpha$ cannot depend on. For example, take a constraint $\alpha[\theta] \sim \beta[\theta']$. Intuitively, we can remove any variable $x$ from the context of $\beta$ if $\theta'$ contains $x \mapsto z$ and $z$ is not present in the image of $\theta$. The second check makes sure $\alpha[\theta]$ does not depend on $z$, since $\theta$ contains substitutions for all free variables that are in scope of $\alpha$. Pruning generalizes this operation such that $\beta$ can appear anywhere in a rigid position of $e$ and verifies that the types in the context of $\beta$ are still well-formed. Finally, for pruning we only consider variable substitutions $\theta$, since only those are potentially invertible when we solve for $\alpha$.

$$
\{C = (\alpha[\theta] \sim e) :: Cs, E = \Delta\} \ \rightsquigarrow \{C = (\alpha[\theta] \sim e') :: Cs, E = \Delta'\} \quad \text{Unif-prune}
$$
where $\Delta = (\Gamma \vdash \alpha : t) :: \ldots$ and $\texttt{prune}(\Delta, \theta, e) \rightsquigarrow \Delta', e'$

---

[4] This extends the possible outcomes of the conversion checker (Section 5).

Unification of identical metavariables handles constraints of shape $\alpha[\theta_1] \sim \alpha[\theta_2]$. If $\theta_1 = [x_1 \mapsto z_1, x_2 \mapsto z_2, x_3 \mapsto y]$ and $\theta_2 = [x_1 \mapsto z_2, x_2 \mapsto z_1, x_3 \mapsto y]$, $\alpha$ can only depend on $y$. To see this, take a potential solution $\alpha = x_1$. Since $x_1$ is mapped to distinct variables, this equation cannot hold. Therefore, $\alpha$ cannot depend on $x_1$. A similar argument applies to $x_2$.

$$\{C = (\alpha[\theta_1] \sim \alpha[\theta_2]) :: Cs, E = \Delta\} \;\rightsquigarrow\; \{C = Cs, E = \Delta'\} \quad \text{Unif-prune-same}$$
where $\Delta = (\Gamma \vdash \alpha : t) :: \Delta_r$ and $\;\mathtt{same}(\Delta, \alpha, \theta_1, \theta_2) \rightsquigarrow \Delta'$

The occurs check verifies two necessary conditions for the constraint to be solvable. The first case in which `occurs` signals an error is if the right-hand side has any free variables in rigid positions not present on the left-hand side: $\mathrm{FV}^{\mathrm{rig}}(e) \not\subset \mathrm{img}(\theta)$. The second case is when the metavariable we are solving for occurs in the potential solution in a strongly rigid position. If this happens, the solution is ill-formed since it would need to be syntactically infinite. Concretely, occurs check signals an error if $\alpha \in \mathrm{FVM}^{\mathrm{srig}}(e)$. Mind that $e \neq \alpha[\theta']$, as this can be simplified via Unif-prune-same.

$$\{C = (\alpha[\theta] \sim e) :: Cs, E = (\Gamma \vdash \alpha : t) :: \Delta\} \;\rightsquigarrow\; \mathbf{fail} \quad \text{Unif-Occurs-rig}$$
where $\mathtt{occurs}(\alpha, \theta, e) \rightsquigarrow \mathbf{fail}$

If the occurs check fails, we postpone the problem until we have a solution for the metavariable containing the offending sub-terms. If the occurs check succeeds, we can try to invert the substitution. If such an inversion exists, we apply it to the right-hand side and obtain a solution.

$$\{C = (\alpha[\theta] \sim e) :: Cs, E = (\Gamma \vdash \alpha : t) :: \Delta\} \rightsquigarrow \qquad\qquad \text{Unif-solve}$$
$$\{C = Cs, E = (\Gamma \vdash \alpha : t := s) :: \Delta\}, \text{ where } s := \mathtt{invert}(\alpha, \theta, e)$$

### 9.4.2 Pruning and Unification of Identical Metavariables

The pruning function $\mathtt{prune}(\Delta, \theta, e)$ traverses the term $e$ recursively, reducing the input to WHNF each step. It outputs a new metavariable state $\Delta'$ and an updated $e'$.

When `prune` traverses the term, it checks variables that are in flexible positions. In particular, when it encounters a meta $\beta[\tau]$ in a rigid position, it tries to prune its arguments. A variable $(x \mapsto t) \in \tau$ is pruned from $\Gamma$, where $\Gamma \vdash \beta : s \in \Delta$, if

- $t$ contains a non-eliminatable [1, p.15] occurrence $y$ that is not in the image of $\theta$. Intuitively, this means that there is no instantiation of $\beta$, such that the offending variable goes away. This means it has to be either a head of a neutral term or have non-eliminatable occurrences in all branches of a comatch.
- $x$ is not used in the types following $x$ in the context $\Gamma$.

If $t$ itself contains $y$ in a flexible position, pruning fails. If both conditions hold, `prune` adds a new metavariable $\Gamma' \vdash \beta' : s$ and a solution $\Gamma \vdash \beta := \beta'[\ldots]$ to $\Delta$. It also substitutes the solution into $e$ and reduces it, if necessary.

Unification of identical metavariables $\mathtt{same}(\Delta, \alpha, \theta_1, \theta_2)$ traverses both substitutions in sync forming a new metavariable $\beta$. If two substitutions map the same argument $x$ to different variables, $\beta$ cannot depend on $x$. If both of them map it to the same variable, same keeps the dependency of $\beta$ on $x$. If the context of $\beta$ is smaller than $\alpha$, same adds $\Gamma \vdash \alpha := \beta[\mathrm{id}_{\Gamma'}]$ and $\Gamma' \vdash \beta$ to $\Delta$, forming $\Delta'$.

### 9.4.3 Occurs Check

Occurs check $\texttt{occurs}(\alpha, \theta, e)$ traverses the term $e$ checking occurrences of $\alpha$ to prevent infinite solutions and whether free variables from $e$ are in the image of $\theta$. It outputs a positive decision **ok**, a negative decision **no**, or a failure **fail**.

We need $\theta$ to be a (potentially non-linear) variable renaming. If we find a substitution $x \mapsto e$, where $e \neq y$, occurs outputs **no**. During the traversal of $e$, if $\texttt{occurs}$ encounters the same metavariable $\alpha$ in a strongly rigid position, it signals a failure **fail**. If it encounters a variable $x$ not in the image of $\theta$, the action depends on the occurrence of the variable. In case of rigid occurrence — signal a **fail**ure, for flexible — output a negative decision **no**. If occurs traverses the term completely without any errors, it outputs a positive decision **ok**.

### 9.4.4 Inversion of Substitutions

Inversion $\texttt{invert}(\alpha, \theta, e)$ computes an application of the inverted substitution $\theta^{-1}$ to $e$. This is not always possible as higher order unification is undecidable in general [35]. We therefore only attempt to solve a subset of problems known as Miller's pattern fragment [52], where a metavariable substitution $\theta$ is a bijective variable renaming. We follow Agda's [4, 5] approach: invert tries to compute the inversion of the substitution $\theta$, checking linearity in the process. If the substitution is not linear, try pruning $\alpha$ with respect to FV($e$). If it does succeed and the inversion of pruned $\theta'$ exists, apply it to $e$ and return the candidate solution.

In order for the solution to be valid two conditions need to hold. For it to be well-scoped, $\theta$ must be invertible on FV($e$). For it to be well-typed, $\theta$ must be invertible on the free variables of the type of every sub-term of $e$ [1, p.11]. This means that we either have to collect type information as we traverse $e$, which would require us to make unification typed and quote back the conversion environment. Alternatively, we could apply the inverted substitution and type-check the solution to make sure that it is well-typed before committing to it.

### 9.5 Discussion

The design of the unification algorithm is guided by the language [10], particularly its symmetric handling of data and codata types. This means that we would like to preserve the well-typedness when a program is re- or defunctionalized. While we can maintain symmetry in the conversion checker, this poses a challenge for unification. In particular, if metavariables stay open for the whole program this might introduce instability of type inference [46, p.10], but there is hope that if the unification algorithm is confluent, we may be able to infer the same solutions. Alternatively, we can freeze metavariables once we exit the scope of a particular definition (not discussed in this paper) and inline the solutions after they have been inferred. In this case we reduce the problem to conversion, which is symmetric, but introduce more syntactical noise in the source program.

We cannot expect the unification algorithm to be terminating since the input is potentially non-terminating and unification relies on reduction. However, for terminating inputs, the algorithm avoids common sources of non-termination found in other unification settings, notably any notion of extensionality for codata [9]. Since we never unfold codata values unless they are projected and have no $\eta$-rules, we avoid that source of non-termination.

While not described in the rules, in the implementation we attempt to solve a constraint without reducing the right-hand side or only reducing partially. If we do find a solution it is a valid one. If the occurs check or the inversion fails, we do need to reduce the terms fully.

There are three classes of simplification rules from prior work [2, fig. 2] we do not implement at the moment:

**Lowering** is used when the type $t$ of the metavariable $\alpha$ gives enough information to refine it. For example, if $t$ reduces to $\Pi(x:A)B$, then we could introduce a new metavariable $\Gamma, x:A \vdash \beta : B$ and solve $\alpha$ with $\lambda x.\beta$. We hypothesize that it can be generalized for any type which uniquely determines its inhabitant. For example, $t := T\sigma$ is a datatype with a single, non-parametric constructor, like $\top$, or the indices $\sigma$ force the choice, like $\mathsf{VNil}(\_) : \mathsf{Vec}(\mathsf{Z}, a)$. Lowering for codata types in our system is further complicated by the fact that each comatch constructor possesses a unique name, which would require metavariables for names [46].

**Decomposition and $\eta$-rules** While we have a decomposition rule (Conv-Dtor) for terms with neutral head, Abel and Pientka describe a more general schema. For functions they simplify $\Gamma \vdash \lambda x.e \sim s$ to $\Gamma, x \vdash e \sim sx$. We do not perform decomposition of codata values, since two comatches are not considered equal even if all their projections are equal, due to the extra condition on label equality. This could be solved by introducing metavariables for labels, as suggested by Liesnikov and Cockx [46]. We do not have $\eta$-contraction for codata types, since the interactions with closure substitutions and the unique labels on comatches make the practicality of such rules unclear. Abel and Pientka also simplify pairs $\Gamma \vdash (e_1, e_2) \sim s$ to $\Gamma \vdash e_1 \sim \mathtt{fst}\ s$ and $\Gamma \vdash e_2 \sim \mathtt{snd}\ s$. In general, if both sides are constructors of an inductive type such a rule is just congruence (Conv-DCtor). If one of the sides is an arbitrary term $e$, we have to know projection-like functions $\mathtt{fst}$ and $\mathtt{snd}$. If they are not unique, then the resulting solution might not be the most general unifier. $\eta$-contraction for inductive types would also need such functions, though not necessarily unique ones.

## 10 Related Work

The algorithmic type system we describe in this article is based on the declarative type system of our prior publication [10]. The main difference is that we make the system algorithmic, add rules for local pattern and copattern matches and describe the elaboration of implicit arguments.

**Tutorial Implementations of Dependent Types** The most accessible tutorial for implementing a typechecker for a dependently typed system is, in our opinion, the excellent book by Friedman and Christiansen [31]. They introduce dependent types from the ground up and develop a description of a simple bi-directional type inference algorithm for dependent types. Löh, McBride, and Swierstra [47], as well as Weirich [75], all describe how to implement a simple dependently typed system in the programming language Haskell, but they do not include metavariables and dependent pattern-matching on user-defined types.

**Self parameters**    As mentioned in Section 1, self parameters are motivated by de- and refunctionalization. A detailed comparison of Polarity with regard to these transformations can be found in previous work [10, p.23]. Several other notions of self parameters have been proposed in the literature. Scala has a concept of self types [58], but they exist in a very different context: Scala builds upon an object calculus with subtyping and has a limited form of dependent types, including path-dependent types [63] and match types [11]. Polarity has full dependent types, but no subtyping. Self types in Scala can refine the type of the object self-reference as long as it stays within the subtyping hierarchy [59]. In Polarity, the self type is refined during dependent pattern matching. In Scala, the self reference (called this by default, though it can be renamed) is always available to method bodies, while in Polarity the self reference is only available on the type-level in the return type of definitions and destructors. Cedille [67, 68] has a form of self-reference using intersection types or $\iota$-types [66]. In contrast to Polarity, Rocq, and others, which add (co)data as primitives, Cedille is based on pure lambda encodings. However, it is possible to encode inductive [66] and coinductive [41] types in Cedille, though the latter does not yet have the "true" coinduction principle.

**Codata Types and Function Types**    Codata types [38] have a long history that we will not recount here in detail. The best introduction to how we understand codata types here can be found in the "Codata in Action" paper [29]. Copattern matching as the term-level construct which inhabits a codata type has been popularized by Abel, Pientka, Thibodeau, and Setzer [3]. It has long been known that codata types subsume function types [64], since they can be expressed as a codata type with one observation. Another dependently-typed system which uses a generic mechanism for specifying inductive and codata types instead of a built-in function type was presented by Basold and Geuvers [8]; our previous work [10, p. 21] contains a more detailed comparison of the systems.

**Dependent Pattern Matching**    Dependent pattern matching was introduced by Coquand [19] as a more convenient way to program with indexed data types. The annotations on pattern matching expressions that indicate the return type are called "motives" and come from work by McBride [50]. The algorithm that we describe in Section 6 is a non-conservative extension over Martin-Löf type theory, since it implies Streicher's Axiom K [65]. If this is undesirable, for example if we work in Homotopy Type Theory, then it is possible to modify the index unification algorithm so that it does not imply Axiom K [13, 14].

**Higher-Order Unification**   We mainly draw from the work of Abel and Pientka [1, 2]. We describe a richer language that includes (co)inductive types, focus on the implementation, and explain the reasoning behind the rules in more detail. We omit $\eta$- and decomposition rules for (co)data types due to the incompatibility with defunctionalization described in Section 9.5. Some of the rules are interesting as future work, but require a more general approach to metavariables [46].

We borrow from the implementation of Agda [5, 57] which employs type-directed conversion and does away with strongly rigid occurrences due to the complexity of Agda with all extensions. We do not need the types and hope to improve error messages by rejecting unsolvable problems earlier. Our work aims to clarify some of the techniques they implement: as far as we know, we are first to describe the pruning for the left-hand side of the equation in a publication.

Unification in Rocq [71, 78, 79] and Lean [24, 56] differs significantly from our algorithm in that it does not always produce a most general unifier (MGU). Due to this fundamental difference in the approach, a direct comparison is challenging. For example, we do not need backtracking in our algorithm and heuristics play a much smaller role. Ziliani and Sozeau [78] include a description of fixpoints, implications regarding advanced features [48], and performance, which we might study in the future. Unlike their work, we cover postponed constraints and coinductive types, though in a somewhat atypical setting. The system implemented by Lean [55] is, in general, closer to the system implemented in Rocq than to the system described in this paper. However, both Rocq and our system implement pruning while Lean does not [78, p.54]. Idris [12] implements pattern [54] unification for a dependently typed calculus with quantities based on the work by Gundry. Idris 2 [70] introduces a lot of new features into the language, including codata. However, the description of the system is not published. The implementation seems similar to Agda's, but weaker, as there is no pruning.

The comparison with Dependent Haskell [30, 36] is mostly interesting with respect to non-termination. We stay on par by avoiding known sources of non-termination and requiring evaluation of potentially unsound proofs. More advanced techniques for unification under binders [37, 73] are currently out of scope for our algorithm as the unification procedure relies on closures instead.

## 11   Future Work

We did not yet prove the soundness of the type inference algorithms and leave this for future work. Another open question is how to combine the metavariable unification with the de- and refunctionalization transformations; we sketched two possible approaches in Section 9.5 but neither of them is fully satisfactory. Finally, in the previous work [10] we describe the missing pieces that have to be worked out if we want to obtain a logically consistent system; we do not repeat them here, but they apply equally to the algorithmic type system described in this article.

## 12    Conclusion

In this article we showed how to typecheck and elaborate a dependently-typed language with features that are often omitted in a first introduction: user-defined data and codata types, dependent pattern matching and implicit arguments. The language we present is unique among dependently-typed programming languages in that it does not presuppose any built-in types: All types have to be declared as either data or codata types. We have therefore provided what we think is a solid foundation for studying the role of polarity in dependently typed languages in the future. The description of the algorithm will also be used as the future reference for the implementation of the Polarity research language, available at polarity-lang.github.io.

**Acknowledgements**    We would like to thank Jesper Cockx for the guidance throughout the work on this paper and Andreas Abel for valuable discussions on the nuances of unification. We are also grateful to Andy Zaidman and Jiří Beneš for their feedback on writing. Finally, we would like to thank András Kovács for his work on elaboration-zoo [42], which aided our initial implementation.

 **List of Notations and Procedures**

Table 3 contains the complete syntax and auxiliary concepts used in this paper and Table 4 lists all the procedures we use.

■ **Table 3** Syntax and notations

---

$$T \in \textsc{TypeNames} \quad K \in \textsc{Constructors} \quad D \in \textsc{Destructors}$$
$$x, y, z \in \textsc{Vars} \quad \alpha, \beta, \gamma \in \textsc{Metavars}$$

| | | | |
|---|---|---|---|
| $e, s, t$ | $::=$ | $x$ | *Variable* |
| | $\mid$ | $e : t$ | *Type annotation* |
| | $\mid$ | $\textbf{let } x : t \coloneqq e; e$ | *Let-binding* |
| | $\mid$ | **Type** | *Universe* |
| | $\mid$ | $T\sigma$ | *Type constructor* |
| | $\mid$ | $K\sigma$ | *Data constructor* |
| | $\mid$ | $e.\textbf{match } x\ \sigma\ \textbf{as } z\ \textbf{return } s\ \{a, \dots\}$ | *Match* |
| | $\mid$ | $e.D\sigma$ | *Destructor* |
| | $\mid$ | $\textbf{comatch } x\ \sigma\ \{o, \dots\}$ | *Comatch* |
| | $\mid$ | $\alpha[\theta]$ | *Contextual metavariables* |
| $a$ | $::=$ | $K\Xi \mapsto e$ | *Case* |
| | $\mid$ | $K\Xi\ \textbf{absurd}$ | *Absurd case* |
| $o$ | $::=$ | $D\Xi \mapsto e$ | *Cocase* |
| | $\mid$ | $D\Xi\ \textbf{absurd}$ | *Absurd cocase* |
| $\Gamma$ | $::=$ | $()\mid \Gamma, x : t\mid \Gamma, x : t \coloneqq e$ | *Typing context* |
| $\Psi, \Xi$ | $::=$ | $()\mid x : e, \Xi$ | *Telescope* |
| $\rho, \sigma$ | $::=$ | $()\mid (e, \sigma)$ | *Telescope substitutions* |
| $\delta$ | $::=$ | $\textbf{data } T\Psi\ \{\,K\Xi : T\rho, \dots\,\}$ | *Data declaration* |
| | $\mid$ | $\textbf{codata } T\Psi\ \{\,(x : T\rho).D\Xi : t, \dots\,\}$ | *Codata declaration* |
| $\Theta$ | $::=$ | $()\mid \Theta, \delta$ | *Global environment* |
| $\phi$ | $::=$ | $\Gamma\mid \phi, x \mapsto e\mid \phi, x \mapsto_c e$ | *Local environment* |
| $\Delta$ | $::=$ | $[\,]\mid (\Gamma \vdash \alpha : t) :: \Delta\mid (\Gamma \vdash \alpha : t \coloneqq e) :: \Delta$ | *Metavariable map* |
| $Cs$ | $::=$ | $\phi \vdash e_1 \sim e_2 :: Cs\mid \phi \vdash \sigma_1 \sim_{\text{arg}} \sigma_2 :: Cs\mid [\,]$ | *Constraints* |
| $\mathscr{U}$ | $::=$ | $\{C = Cs, E = \Delta\}$ | *Unification context* |
| $\theta, \eta$ | $::=$ | $()\mid \theta, x \mapsto t$ | *Context substitutions* |

---

 **Well-formedness of Programs**

At the outermost level we check that global environments are well-formed, which we write as $\textsc{Wf}(\Theta)$. A global environment is well-formed if all of its declarations are

■ **Table 4** List of Procedures

| Syntax | Meaning | Reference |
|--------|---------|-----------|
| $e[\phi] \triangleright e'[\phi']$ | Single-step reduction towards WHNF | Section 4 |
| $e[\phi] \triangleright^* w[\phi']$ | Multi-step reduction to WHNF | Section 4 |
| $e \rightsquigarrow w$ | Normalize to WHNF and quote back | Section 4 |
| $\mathtt{conv}(\Gamma, e_1, e_2)$ | Conversion check | Section 5 |
| $\mathtt{unify}_\iota(\rho_1, \rho_2)$ | Index Unification | Section 6 |
| $\mathtt{register}(\Delta, \alpha, t, \Gamma)$ | Register a metavariable | Section 8 |
| $\mathtt{prune}(\Delta, \theta, e)$ | Pruning | Section 9 |
| $\mathtt{same}(\Delta, \alpha, \theta_1, \theta_2)$ | Unification of identical metavariables | Section 9 |
| $\mathtt{occurs}(\alpha, \theta, e)$ | Occurs check | Section 9 |
| $\mathtt{invert}(\theta, e)$ | Inversion of delayed substitutions | Section 9 |

well-formed, and we check each declaration in the context of all the declarations that have been introduced before.[5] We write $\textsc{Wf}(\Theta \vdash \delta)$ to express that the declaration $\delta$ is well-formed in the global environment $\Theta$.

$$\frac{}{\textsc{Wf}(())} \text{ Wf-Empty} \qquad \frac{\textsc{Wf}(\Theta) \qquad \textsc{Wf}(\Theta, \delta \vdash \delta)}{\textsc{Wf}(\Theta, \delta)} \text{ Wf-Cons}$$

There are two kinds of declarations to consider: data declarations and codata declarations. In the wellformedness rule for data types we check that the indices are well-formed ❶, that the arguments of every data constructor are well-formed ❷ and that for every data constructor the type constructor receives valid arguments ❸.

$$\frac{\Theta \vdash \Psi \text{ tele ❶} \qquad \forall i : [\Theta \vdash \Xi_i \text{ tele ❷} \qquad \Theta \mid \Xi_i \vdash \rho_i \Leftarrow \Psi \text{ ❸}]}{\textsc{Wf}(\Theta \vdash \text{data } T \Psi \, \{ \, K\Xi : T\rho, \dots \})} \text{ Wf-Data}$$

For codata types we additionally check the return type of each destructor in the context extended with the term on which we apply the destructor ❶.

$$\frac{\Theta \vdash \Psi \text{ tele}}{\frac{\forall i : [\Theta \vdash \Xi_i \text{ tele} \quad \Theta \mid \Xi_i \vdash \rho_i \Leftarrow \Psi \quad \Theta \mid \Xi_i, x : T\rho \vdash t \Leftarrow \textbf{Type ❶}]}{\textsc{Wf}(\Theta \vdash \text{codata } T\Psi \, \{ \, (x : T\rho).D\Xi : t, \dots \})}} \text{ Wf-Codata}$$

We currently do not check positivity of (co)data types which is a well-known source of non-termination. However, consistency is not a concern for the system presented here; we also use **Type** : **Type**, and have no termination or productivity checks.

## <span style="background-color:blue;color:white">C</span> Expressiveness of Self Parameters

In this section we provide an example which, we believe cannot be expressed without self parameters in codata types. Using an encoding proposed by Fu and Stump [32], we can represent a natural number using its induction principle. The induction principle for the number n constructs, for any property P: Nat -> Type, a proof of P(n) by applying the induction step n-times to the proof of the base case P(Z).

---

[5] This is a simplification over the implementation of Polarity which allows mutually recursive declarations, a requirement for de- and refunctionalization to always be possible.

```
1  codata Nat {
2    (n: Nat).ind(P: Nat -> Type,
3              base: P.ap(a:=Nat, b:=Type, Z),
4              step: Pi(Nat, StepFun(P))
5              )
6      : P.ap(a:=Nat, b:=Type, n),
7  }
8  codef Z: Nat { .ind(P, base, step) => base }
9  codef S(m: Nat): Nat {
10   .ind(P, base, step) =>
11     step.dap(Nat, StepFun(P), m)
12       .ap(a:=P.ap(a:=Nat, b:=Type, m),
13         b:=P.ap(a:=Nat, b:=Type, S(m)),
14         m.ind(P, base, step))
15 }
```

```
1  data Nat { S(m: Nat), Z }
2  def (n: Nat).ind(P: Nat -> Type,
3             base: P.ap(a:=Nat, b:=Type, Z),
4             step: Pi(Nat, StepFun(P))
5             )
6     : P.ap(a:=Nat, b:=Type, n) {
7    S(m) =>
8      step.dap(Nat, StepFun(P), m)
9        .ap(a:=P.ap(a:=Nat, b:=Type, m),
10         b:=P.ap(a:=Nat, b:=Type, S(m)),
11         m.ind(P, base, step)),
12   Z => base,
13 }
```

On the left-hand side, we define the Fu-Stump encoding of natural numbers in Polarity. The defunctionalized version on the right-hand side shows how the Fu-Stump encoding on the left corresponds to a program which defines an induction principle on Peano natural numbers. In both representations, StepFun abbreviates a function type that takes a proof of P(x) to a proof of P(S(x)).

```
1  codef StepFun(P: Nat -> Type): Fun(Nat, Type) {
2    .ap(_, _, x) => P.ap(a:=Nat, b:=Type, x) -> P.ap(a:=Nat, b:=Type, S(x))
3  }
```

## D  Formal Definition of Occurrences

Occurrences $o^{\{\text{srig,rig,flx}\}}$ are defined on normal forms $v$. In the definitions of $o$ below, the dot $\cdot$ stands for an occurrence.

**Definition 7** (Neutral Terms and Normal Forms).

$$n^v ::= x \mid n^v.D\sigma^v \mid n^v.\textbf{match } x \; \sigma^v \textbf{ as } z \textbf{ return } s^v \; \{a, \dots\}$$
$$v ::= n^v \mid \textbf{\textit{Type}} \mid T\sigma^v \mid K\sigma^v \mid \textbf{comatch } x \; \sigma^v \; \{o, \dots\} \qquad \text{Normal Form}$$

**Definition 8** (Strongly rigid occurrences).

$$o^{srig} ::= \cdot \mid K\sigma^{srig} \mid \textbf{comatch } x \; \sigma^{srig} \; \{\dots\}$$
$$\sigma^{srig} ::= (o^{srig}, \sigma^v) \mid (v, \sigma^{srig})$$

We only present the grammar for rigid occurrences, since only they are used in the unification algorithm. To obtain the grammar for weakly rigid occurrences, the reader has to factor out the cases of strongly rigid occurrences from $o^{rig}$.

**Definition 9** (Rigid occurrences).

$$
\begin{aligned}
o^{prig} ::= &\; \cdot \mid n^v.D\sigma^{rig} \mid o^{prig}.D\sigma^v \\
\mid &\; n^v.\textbf{match } x \; \sigma^{rig} \textbf{ as } z \textbf{ return } s \; \{\dots\} \mid n^v.\textbf{match } x \; \sigma \textbf{ as } z \textbf{ return } o^{rig}\{\dots\} \\
\mid &\; o^{prig}.\textbf{match } x \; \sigma \textbf{ as } z \textbf{ return } s \; \{\dots\} \\
o^{rig} ::= &\; o^{prig} \mid K\sigma^{rig} \mid \textbf{comatch } x \; \sigma^{rig} \; \{D\Xi \mapsto v, \dots\} \\
\sigma^{rig} ::= &\; (o^{rig}, \sigma^v) \mid (v, \sigma^{rig})
\end{aligned}
$$

**Definition 10** (Flexible occurrences).

$$
\begin{aligned}
o^{rflx} \quad &::= \quad n^{v}.D\sigma^{qflx} \mid o^{rflx}.D\sigma^{v} \\
&\mid \quad n^{v}.\textbf{match } x\ \sigma^{qflx} \textbf{ as } z \textbf{ return } s\ \{\ldots\} \mid o^{rflx}.\textbf{match } x\ \sigma^{v} \textbf{ as } z \textbf{ return } s\ \{\ldots\} \\
o^{qflx} \quad &::= \quad \cdot \mid o^{rflx} \mid K\sigma^{qflx} \mid \textbf{comatch } x\ \sigma^{qflx}\ \{\ldots\} \\
\sigma^{qflx} \quad &::= \quad (o^{qflx}, \sigma^{v}) \mid (v, \sigma^{qflx}) \\
o^{pflx} \quad &::= \quad n^{v}.D\sigma^{flx} \mid o^{pflx}.D\sigma^{v} \\
&\mid \quad n^{v}.\textbf{match } x\ \sigma^{v} \textbf{ as } z \textbf{ return } s\ \{\ldots\} \mid n^{v}.\textbf{match } x\ \sigma^{v} \textbf{ as } z \textbf{ return } o^{flx}\{\ldots\} \\
&\mid \quad o^{pflx}.\textbf{match } x\ \sigma^{v} \textbf{ as } z \textbf{ return } s\ \{\ldots\} \\
o^{flx} \quad &::= \quad \alpha[x \mapsto o^{qflx}, \ldots] \mid K\sigma^{flx} \mid \textbf{comatch } x\ \sigma^{flx}\ \{\ldots\} \\
\sigma^{flx} \quad &::= \quad (o^{flx}, \sigma^{v}) \mid (v, \sigma^{flx})
\end{aligned}
$$

**Examples**  The following examples illustrate the differences between strongly rigid, weakly rigid, and flexible occurrences.

- Strongly rigid occurrences.
  - **inductive constructors**. Equations like $\alpha = K(\alpha)$ could be accepted by our system for now due to the lack of termination checking, in particular using a recursive let binding. Without a recursive let the inductive constructor fixes the head of the term, meaning that a solution would be of shape $\alpha = K(K(K(\ldots)),$ which is not allowed. We reject such equations in any case to keep the system consistent with a future termination check.
  - **comatch closure**. A comatch closure can be strongly rigid to maintain symmetry with inductive constructors, even if a solution might exist. This means that an equation like $\alpha = \textbf{comatch } x\ s\ \{D\,y \mapsto \alpha\}$ is solvable, since $\alpha$ can be instantiated to $x$. While an equation like $\alpha = \textbf{comatch } x\ [z \mapsto \alpha]\{D\Xi \mapsto z\}$ would not be solvable, despite being superficially similar.

- Weakly rigid occurrences.
  - **head of a projection**. A variable being projected from is weakly rigid, as its structure can be instantiated to satisfy the projection. E.g., $\alpha = \alpha.\textrm{tl}$ is solvable with $\alpha = \textbf{comatch } x\ \{tl \mapsto x\}$. This contradicts a naive generalization of a definition from Abel and Pientka [2, p.13], following which $\alpha$ should be strongly rigid as is not in the evaluation context of a free variable. The original definition still seems to not account for cases like $\alpha = (\textrm{succ}(\alpha.snd), 0)$. This would be considered strongly rigid, but it has a solution $\alpha = (1, 0)$. In practice, however, the definition is fine as used in the original paper, since any metavariable of this form would be decomposed into two, breaking down the equation into $\alpha = \textrm{succ}(w)$ and $w = 0$.
  - **argument to a projection**. An occurrence within an evaluation context of a free variable is weakly rigid, allowing for solutions that specialize the variable's context. An equation $\alpha[x \mapsto z] = z.D\alpha[x \mapsto \textbf{comatch}\{D\Xi \mapsto \textrm{zero}\}]$ is solvable with $\alpha = z.D(\textrm{zero})$. This also follows from the original definition [2], since the occurrence is in an evaluation context of a free variable.
  - **return type of a match**. $\alpha = n.\textbf{match } x\ \sigma \textbf{ as } x \textbf{ return } \alpha\ \{\ldots\}$
    While we currently lack a concrete counterexample for an equation like this, the

scenario raises concerns about the potential for circular type dependencies. If a counterexample is found, this could become a strongly rigid occurrence. For now we consider it weakly rigid as a conservative approximation.

- **scrutinee of a match**. A meta-variable as the scrutinee of a match is weakly rigid because its instantiation can enable the match to compute further. For instance, $\alpha[z \mapsto x] = (\alpha[z \mapsto \text{succ } x]).\textbf{match} \{\text{succ } y \mapsto y\}$ does have a solution, namely $\alpha = z$. This is another counterexample to a naive generalization of strongly rigid variables.

▪ Flexible occurrences.

- **in a delayed substitution** of a meta-variable. Flexible occurrences appear in the delayed substitution of a meta-variable, representing the most pliable form.

### References

[1] Andreas Abel and Brigitte Pientka. "Extensions to Miller's Pattern Unification for Dependent Types and Records". Under consideration for publication in Math. Struct. in Comp. Science. May 2, 2018. URL: https://www.cs.mcgill.ca/~b pientka/papers/unif_miller60.pdf (visited on 2025-10-16).

[2] Andreas Abel and Brigitte Pientka. "Higher-Order Dynamic Pattern Unification for Dependent Types and Records". In: *Typed Lambda Calculi and Applications*. Edited by Luke Ong. Berlin, Heidelberg: Springer Berlin Heidelberg, 2011, pages 10–26. ISBN: 978-3-642-21691-6. DOI: 10.1007/978-3-642-21691-6_5.

[3] Andreas Abel, Brigitte Pientka, David Thibodeau, and Anton Setzer. "Copatterns: Programming Infinite Structures by Observations". In: *Proceedings of the 40th Annual ACM SIGPLAN-SIGACT Symposium on Principles of Programming Languages*. POPL '13. Rome, Italy: Association for Computing Machinery, 2013, pages 27–38. ISBN: 9781450318327. DOI: 10.1145/2480359.2429075.

[4] Agda Developers. *./src/full/Agda/TypeChecking/MetaVars.hs at Release-2.8.0 · Agda/Agda*. URL: https://github.com/agda/agda/blob/release-2.8.0/src/full/Ag da/TypeChecking/MetaVars.hs/#L1032-L1087 (visited on 2025-08-01).

[5] Agda Developers. *Agda*. Version 2.8.0. June 2025. URL: https://agda.readthedo cs.io/en/v2.8.0/ (visited on 2025-07-01).

[6] Hendrik Pieter Barendregt. *The Lambda Calculus: Its Syntax and Semantics*. New York, NY, USA: Elsevier, 1981. ISBN: 978-1848900660.

[7] Gilles Barthe, Venanzio Capretta, and Olivier Pons. "Setoids in Type Theory". In: *Journal of Functional Programming* 13.2 (2003), pages 261–293. ISSN: 1469-7653, 0956-7968. DOI: 10.1017/S0956796802004501.

[8] Henning Basold and Herman Geuvers. "Type Theory Based on Dependent Inductive and Coinductive Types". In: *Proceedings of the Symposium on Logic in Computer Science*. New York: Association for Computing Machinery, July 2016, pages 327–336. ISBN: 9781450343916. DOI: 10.1145/2933575.2934514.

[9]   Ulrich Berger and Anton Setzer. "Undecidability of Equality for Codata Types".
      In: *Coalgebraic Methods in Computer Science*. Edited by Corina Cîrstea. Springer.
      2018, pages 34–55. DOI: 10.1007/978-3-030-00389-0_4.

[10]  David Binder, Ingo Skupin, Tim Süberkrüb, and Klaus Ostermann. "Deriving
      Dependently-Typed OOP from First Principles". In: *Proc. ACM Program. Lang.*
      8.OOPSLA1 (Apr. 2024). DOI: 10.1145/3649846.

[11]  Olivier Blanvillain, Jonathan Immanuel Brachthäuser, Maxime Kjaer, and Mar-
      tin Odersky. "Type-level Programming with Match Types". In: *Proceedings of
      the ACM on Programming Languages* 6.POPL (2022), pages 1–24. DOI: 10.1145/3
      498698.

[12]  Edwin Brady. "Idris, a General-Purpose Dependently Typed Programming
      Language: Design and Implementation". In: *Journal of Functional Programming*
      23.5 (2013), pages 552–593. DOI: 10.1017/S095679681300018X.

[13]  Jesper Cockx. "Dependent Pattern Matching and Proof-Relevant Unification".
      PhD thesis. KU Leuven, 2017. URL: https://lirias.kuleuven.be/1656778.

[14]  Jesper Cockx, Dominique Devriese, and Frank Piessens. "Pattern Matching
      without K". In: *International Conference on Functional Programming*. New York,
      NY, USA: Association for Computing Machinery, 2014, pages 257–268. DOI:
      10.1145/2628136.2628139.

[15]  Jesper Cockx, Dominique Devriese, and Frank Piessens. "Unifiers as Equiva-
      lences: Proof-Relevant Unification of Dependently Typed Data". In: *SIGPLAN
      Not.* 51.9 (Sept. 2016), pages 270–283. ISSN: 0362-1340. DOI: 10.1145/3022670
      .2951917.

[16]  William R. Cook. "On Understanding Data Abstraction, Revisited". In: *Proceed-
      ings of the Conference on Object-Oriented Programming, Systems, Languages and
      Applications: Onward! Essays*. Orlando: Association for Computing Machinery,
      2009, pages 557–572. DOI: 10.1145/1640089.1640133.

[17]  Thierry Coquand. "An Algorithm for Testing Conversion in Type Theory". In:
      *Logical Frameworks*. USA: Cambridge University Press, Oct. 1, 1991, pages 255–
      279. ISBN: 978-0-521-41300-8. DOI: 10.1017/CBO9780511569807.011.

[18]  Thierry Coquand. "Infinite Objects in Type Theory". In: *Types for Proofs and
      Programs*. Edited by Henk Barendregt and Tobias Nipkow. Redacted by Gerhard
      Goos and Juris Hartmanis. Volume 806. Berlin, Heidelberg: Springer, 1994,
      pages 62–78. ISBN: 978-3-540-58085-0. DOI: 10.1007/3-540-58085-9_72.

[19]  Thierry Coquand. "Pattern Matching With Dependent Types". In: *Proceedings of
      the 1992 Workshop on Types for Proofs and Programs*. Edited by Bengt Nordström,
      Kent Pettersson, and Gordon Plotkin. Bastad, Sweden, 1992, pages 66–79.

[20]  Olivier Danvy and Kevin Millikin. "Refunctionalization at Work". In: *Science of
      Computer Programming* 74.8 (2009), pages 534–549. DOI: 10.1016/j.scico.2007.1
      0.007.

[21]    Olivier Danvy and Lasse R. Nielsen. "Defunctionalization at Work". In: *Proceedings of the Conference on Principles and Practice of Declarative Programming*. Florence, 2001, pages 162–174. ISBN: 1-58113-388-X. DOI: 10.1145/773184.773202.

[22]    Nicolaas G. de Bruijn. "Lambda Calculus Notation with Nameless Dummies, a Tool for Automatic Formula Manipulation, with Application to the Church-Rosser Theorem". In: *Indagationes mathematicae*. Volume 75. 5. Elsevier. 1972, pages 381–392. DOI: 10.1016/1385-7258(72)90034-0.

[23]    Nicolaas G. de Bruijn. "Telescopic Mappings in Typed Lambda Calculus". In: *Information and Computation* 91.2 (1991), pages 189–204. ISSN: 08905401. DOI: 10.1016/0890-5401(91)90066-B.

[24]    Leonardo de Moura, Soonho Kong, Jeremy Avigad, Floris van Doorn, and Jakob von Raumer. "The Lean Theorem Prover (System Description)". In: *Automated Deduction - CADE-25*. Edited by Amy P. Felty and Aart Middeldorp. Lecture Notes in Computer Science. Cham: Springer International Publishing, 2015, pages 378–388. ISBN: 978-3-319-21401-6. DOI: 10.1007/978-3-319-21401-6_26.

[25]    Gilles Dowek, Thérèse Hardin, Claude Kirchner, and Frank Pfenning. "Unification via Explicit Substitutions: The Case of Higher-Order Patterns". In: *Logic Programming, Proceedings of the 1996 Joint International Conference and Symposium on Logic Programming, Bonn, Germany, September 2-6, 1996*. Edited by Michael J. Maher. MIT Press, 1996, pages 259–273. ISBN: 9780262291309. URL: https://ieeexplore.ieee.org/xpl/articleDetails.jsp?arnumber=6278922.

[26]    Paul Downen and Zena M. Ariola. "Beyond Polarity: Towards a Multi-Discipline Intermediate Language with Sharing". In: *27th EACSL Annual Conference on Computer Science Logic (CSL 2018)*. Edited by Dan Ghica and Achim Jung. Volume 119. Leibniz International Proceedings in Informatics (LIPIcs). Dagstuhl, Germany: Schloss Dagstuhl - Leibniz-Zentrum für Informatik, 2018, 21:1–21:23. ISBN: 978-3-95977-088-0. DOI: 10.4230/LIPIcs.CSL.2018.21.

[27]    Paul Downen and Zena M. Ariola. "Compiling With Classical Connectives". In: *Logical Methods in Computer Science* Volume 16, Issue 3 (Aug. 2020). DOI: 10.23638/LMCS-16(3:13)2020.

[28]    Paul Downen, Philip Johnson-Freyd, and Zena M. Ariola. "Structures for Structural Recursion". In: *Proceedings of the 20th ACM SIGPLAN International Conference on Functional Programming*. ICFP 2015. Vancouver, BC, Canada: Association for Computing Machinery, 2015, pages 127–139. ISBN: 9781450336697. DOI: 10.1145/2784731.2784762.

[29]    Paul Downen, Zachary Sullivan, Zena M. Ariola, and Simon Peyton Jones. "Codata in Action". In: *European Symposium on Programming*. ESOP '19. Springer. 2019, pages 119–146. DOI: 10.1007/978-3-030-17184-1_5.

[30]    Richard Eisenberg. "Dependent Types in Haskell: Theory and Practice". PhD thesis. University of Pennsylvania, 2016. DOI: 20.500.14332/29166.

[31] Daniel P. Friedman and David Thrane Christiansen. *The Little Typer*. MIT Press, 2018. ISBN: 9780262536431.

[32] Peng Fu and Aaron Stump. "Self Types for Dependently Typed Lambda Encodings". In: *International Conference on Rewriting Techniques and Applications*. Edited by Gilles Dowek. Springer. 2014, pages 224–239. ISBN: 978-3-319-08918-8. DOI: 10.1007/978-3-319-08918-8_16.

[33] Erich Gamma, Richard Helm, Ralph Johnson, and John Vlissides. *Design Patterns: Elements of Reusable Object-Oriented Software*. Boston: Addison-Wesley Publishing Co., 1995. ISBN: 3827330432.

[34] Jean-Yves Girard. "Interprétation fonctionelle et élimination des coupures de l'arithmétique d'ordre supérieur". Thése de Doctorat d'Etat. Université de Paris VII, 1972. OCLC: 65409853. URL: https://girard.perso.math.cnrs.fr/These.pdf (visited on 2025-10-16).

[35] Warren D. Goldfarb. "The Undecidability of the Second-Order Unification Problem". In: *Theoretical Computer Science* 13.2 (Jan. 1, 1981), pages 225–230. ISSN: 0304-3975. DOI: 10.1016/0304-3975(81)90040-2.

[36] Adam Gundry. "Type Inference, Haskell and Dependent Types". PhD thesis. University of Strathclyde, 2013. DOI: 10.48730/jt3g-ws74.

[37] Adam Gundry and Conor McBride. "A Tutorial Implementation of Dynamic Pattern Unification". Unpublished Article. 2013. URL: https://adam.gundry.co.uk/pub/pattern-unify/ (visited on 2025-10-16).

[38] Tatsuya Hagino. "Codatatypes in ML". In: *Journal of Symbolic Computation* 8.6 (1989), pages 629–650. DOI: 10.1016/S0747-7171(89)80065-3.

[39] Yulong Huang and Jeremy Yallop. "Defunctionalization with Dependent Types". In: *Proceedings of the ACM on Programming Languages* 7.PLDI (June 2023). DOI: 10.1145/3591241.

[40] Antonius J. C. Hurkens. "A Simplification of Girard's Paradox". In: *Proceedings of the Conference on Typed Lambda Calculi and Applications*. London: Springer, 1995, pages 266–278. ISBN: 3-540-59048-X. DOI: 10.1007/BFb0014058.

[41] Christopher Jenkins, Aaron Stump, and Larry Diehl. "Efficient Lambda Encodings for Mendler-style Coinductive Types in Cedille". In: *Proceedings Eighth Workshop on Mathematically Structured Functional Programming, MSFP@ETAPS 2020, Dublin, Ireland, 25th April 2020*. Edited by Max S. New and Sam Lindley. Volume 317. EPTCS. 2020, pages 72–97. DOI: 10.4204/EPTCS.317.5.

[42] András Kovács. *Elaboration-Zoo*. Mar. 2025. URL: https://github.com/AndrasKovacs/elaboration-zoo (visited on 2025-03-01).

[43] M. M. Lehman. "On Understanding Laws, Evolution, and Conservation in the Large-Program Life Cycle". In: *Journal of Systems and Software* 1 (1979), 213–221. ISSN: 0164-1212. DOI: 10.1016/0164-1212(79)90022-0.

[44] Zengyang Li, Paris Avgeriou, and Peng Liang. "A Systematic Mapping Study on Technical Debt and Its Management". In: *Journal of Systems and Software* 101 (2015), pages 193–220. ISSN: 0164-1212. DOI: 10.1016/j.jss.2014.12.027.

[45]   B. P. Lientz, E. B. Swanson, and G. E. Tompkins. "Characteristics of Application Software Maintenance". In: *Communications of the ACM* 21.6 (1978), pages 466–471. ISSN: 0001-0782. DOI: 10.1145/359511.359522.

[46]   Bohdan Liesnikov and Jesper Cockx. "ExEl: Building an Elaborator Using Extensible Constraints". In: *Proceedings of the 35th Symposium on Implementation and Application of Functional Languages*. IFL '23. Braga, Portugal: Association for Computing Machinery, 2024. ISBN: 9798400716317. DOI: 10.1145/3652561.3 652565.

[47]   Andres Löh, Conor McBride, and Wouter Swierstra. "A Tutorial Implementation of a Dependently Typed Lambda Calculus". In: *Fundamenta Informaticae* 102.2 (Apr. 2010), pages 177–207. ISSN: 0169-2968. DOI: 10.3233/FI-2010-304.

[48]   Assia Mahboubi and Enrico Tassi. "Canonical Structures for the Working Coq User". In: *Proceedings of the 4th International Conference on Interactive Theorem Proving*. ITP'13. Berlin, Heidelberg: Springer-Verlag, 2013, pages 19–34. ISBN: 978-3-642-39633-5. DOI: 10.1007/978-3-642-39634-2_5.

[49]   Conor McBride. *Basics of Bidirectionalism*. pigworker in a space. Aug. 6, 2018. URL: https://pigworker.wordpress.com/2018/08/06/basics-of-bidirectionalism/ (visited on 2025-06-02).

[50]   Conor McBride. "Elimination with a Motive". In: *Types for Proofs and Programs*. Edited by Paul Callaghan, Zhaohui Luo, James McKinna, Robert Pollack, and Robert Pollack. Volume 2277. Lecture Notes in Computer Science. Berlin, Heidelberg: Springer, 2002, pages 197–216. DOI: 10.1007/3-540-45842-5_13.

[51]   Conor McBride and James McKinna. "The View from the Left". In: *Journal of Functional Programming* 14.1 (2004), pages 69–111. ISSN: 0956-7968. DOI: 10.1017/S0956796803004829.

[52]   Dale Miller. "A Logic Programming Language with Lambda-Abstraction, Function Variables, and Simple Unification". In: *Journal of Logic and Computation* 1.4 (1991), pages 497–536. DOI: 10.1093/logcom/1.4.497.

[53]   Dale Miller. "Unification of Simply Typed Lambda-Terms as Logic Programming". In: *Logic Programming: Proceedings of the Eighth International Conference*. Edited by Koichi Furukawa. MIT Press, 1991. URL: https://repository.upenn.ed u/handle/20.500.14332/7376.

[54]   Dale Miller. "Unification under a Mixed Prefix". In: *Journal of Symbolic Computation* 14.4 (1992), pages 321–358. ISSN: 0747-7171. DOI: 10.1016/0747-7171(92 )90011-R.

[55]   Leonardo de Moura, Jeremy Avigad, Soonho Kong, and Cody Roux. *Elaboration in Dependent Type Theory*. Version 2. Dec. 17, 2015. arXiv: 1505.04324 [cs.LO]. Pre-published.

[56]  Leonardo de Moura and Sebastian Ullrich. "The Lean 4 Theorem Prover and Programming Language". In: *Automated Deduction – CADE 28*. Edited by André Platzer and Geoff Sutcliffe. Lecture Notes in Computer Science. Cham: Springer International Publishing, 2021, pages 625–635. ISBN: 978-3-030-79876-5. DOI: 10.1007/978-3-030-79876-5_37.

[57]  Ulf Norell. "Towards a Practical Programming Language Based on Dependent Type Theory". PhD thesis. Göteborg, Sweden: Chalmers University of Technology and Göteborg University, 2007. 166 pages. ISBN: 978-91-7291-996-9. URL: https://research.chalmers.se/en/publication/46311.

[58]  Martin Odersky, Vincent Cremet, Christine Röckl, and Matthias Zenger. "A Nominal Theory of Objects with Dependent Types". In: *European Conference on Object-Oriented Programming*. Springer. 2003, pages 201–224. DOI: 10.1007/978-3-540-45070-2_10.

[59]  Martin Odersky and Matthias Zenger. "Scalable Component Abstractions". In: *Proceedings of the 20th annual ACM SIGPLAN conference on Object-oriented programming, systems, languages, and applications*. 2005, pages 41–57. DOI: 10.1145/1094811.1094815.

[60]  Pierre-Marie Pédrot. *Coq Should Have Subject Reduction ([Cbv] Should Not Produce Ill-Typed Terms When Cofixpoints Are Involved) · Issue #5288 · Rocq-Prover/Rocq*. Dec. 2016. URL: https://github.com/rocq-prover/rocq/issues/5288 (visited on 2025-05-01).

[61]  Pierre-Marie Pédrot. *Deprecate Positive Coinductive Types · Pull Request #7536 · Rocq-Prover/Rocq*. May 2018. URL: https://github.com/rocq-prover/rocq/pull/7536 (visited on 2025-05-01).

[62]  Jason Reed. "Higher-Order Constraint Simplification in Dependent Type Theory". In: *Proceedings of the Fourth International Workshop on Logical Frameworks and Meta-Languages: Theory and Practice*. LFMTP '09. New York, NY, USA: Association for Computing Machinery, Aug. 2, 2009, pages 49–56. ISBN: 978-1-60558-529-1. DOI: 10.1145/1577824.1577832.

[63]  Tiark Rompf and Nada Amin. "Type Soundness for Dependent Object Types (DOT)". In: *Proceedings of the 2016 ACM SIGPLAN International Conference on Object-Oriented Programming, Systems, Languages, and Applications*. OOPSLA 2016. New York, NY, USA: Association for Computing Machinery, Oct. 19, 2016, pages 624–641. ISBN: 978-1-4503-4444-9. DOI: 10.1145/2983990.2984008.

[64]  Anton Setzer. "Java as a Functional Programming Language". In: *Types for Proofs and Programs*. Edited by Herman Geuvers and Freek Wiedijk. Springer. Berlin, Heidelberg, 2003, pages 279–298. DOI: 10.1007/3-540-39185-1_16.

[65]  Thomas Streicher. *Investigations into Intensional Type Theory*. Habilitationsschrift, Ludwig-Maximilians-Universität München. 1993. URL: https://www2.mathematik.tu-darmstadt.de/~streicher/HabilStreicher.pdf (visited on 2025-10-20).

[66] Aaron Stump. "From Realizability to Induction via Dependent Intersection". In: *Annals of Pure and Applied Logic* 169.7 (July 1, 2018), pages 637–655. ISSN: 0168-0072. DOI: 10.1016/j.apal.2018.03.002.

[67] Aaron Stump. "The Calculus of Dependent Lambda Eliminations". In: *Journal of Functional Programming* 27 (2017), e14. DOI: 10.1017/S0956796817000053.

[68] Aaron Stump and Christopher Jenkins. *Syntax and Semantics of Cedille*. 2021. arXiv: 1806.04709 [cs.PL]. Pre-published.

[69] Neil Tennant. "Proof and Paradox". In: *Dialectica* 36.2-3 (1982), pages 265–296. DOI: 10.1111/j.1746-8361.1982.tb00820.x.

[70] The Idris Community. *Documentation for the Idris 2 Language*. URL: https://idris2.readthedocs.io/en/latest/ (visited on 2025-05-01).

[71] The Rocq Development Team. *The Rocq Prover*. Version 9.0.0. Mar. 12, 2025. URL: https://rocq-prover.org/doc/V9.0.0/refman/index.html (visited on 2025-06-01).

[72] David Thibodeau, Andrew Cave, and Brigitte Pientka. "Indexed Codata Types". In: *Proceedings of the International Conference on Functional Programming*. ICFP 2016. Nara, Japan: Association for Computing Machinery, 2016, pages 351–363. ISBN: 978-1-4503-4219-3. DOI: 10.1145/2951913.2951929.

[73] Víctor López Juan. "Practical Heterogeneous Unification for Dependent Type Checking". Chalmers University of Technology, 2021. ISBN: 9789179055837. URL: https://research.chalmers.se/en/publication/527051.

[74] Philip Wadler. "The Expression Problem". Note to Java Genericity mailing list. Nov. 1998. URL: https://homepages.inf.ed.ac.uk/wadler/papers/expression/expression.txt.

[75] Stephanie Weirich. *Implementing Dependent Types in pi-forall*. 2023. arXiv: 2207.02129 [cs.PL].

[76] Brendan Zabarauskas. *Add Example of Set Objects · Pull Request #506 · Polarity-Lang/Polarity*. Mar. 2025. URL: https://github.com/polarity-lang/polarity/pull/506 (visited on 2025-05-01).

[77] Noam Zeilberger. "On the Unity of Duality". In: *Annals of Pure and Applied Logic* 153.1-3 (2008), pages 66–96. DOI: 10.1016/j.apal.2008.01.001.

[78] Beta Ziliani and Matthieu Sozeau. "A Comprehensible Guide to a New Unifier for CIC Including Universe Polymorphism and Overloading". In: *Journal of Functional Programming* 27 (2017). ISSN: 0956-7968, 1469-7653. DOI: 10.1017/S0956796817000028.

[79] Beta Ziliani and Matthieu Sozeau. "A Unification Algorithm for Coq Featuring Universe Polymorphism and Overloading". In: *Proceedings of the 20th ACM SIGPLAN International Conference on Functional Programming*. ICFP 2015. New York, NY, USA: Association for Computing Machinery, Aug. 29, 2015, pages 179–191. ISBN: 978-1-4503-3669-7. DOI: 10.1145/2784731.2784751.

## About the authors

**Bohdan Liesnikov** is finishing a PhD at TU Delft. Their research focuses on type checkers and elaborators for dependently typed languages. Bohdan explores how different algorithms and design choices impact the end user. The goal is to develop implementation principles for more predictable, helpful, and ultimately more empowering tools for programmers. Contact Bohdan at B.Liesnikov@tudelft.nl or via website bohdan.liesnikov.name.
https://orcid.org/0009-0000-2216-8830

**David Binder** is a type theorist interested in applications of duality to the theory and implementation of programming languages. This includes, in particular, dualities arising from the difference between data and codata types, subtyping, proof and refutation in the sequent calculus and linear logic. Contact David at D.Binder@kent.ac.uk.
https://orcid.org/0000-0003-1272-0972

**Tim Süberkrüb** is one of the authors of the *Polarity* research programming language. He is interested in further exploring the design space of dependently typed languages, both in theory and practice. Contact Tim at tim.sueberkrueb@uni-tuebingen.de
https://orcid.org/0000-0001-8709-6321